\documentclass[12pt]{spieman}  
\usepackage{amsmath,amsfonts,amssymb}
\usepackage{graphicx}
\usepackage{subcaption}
\usepackage{setspace}
\usepackage{tocloft}
\usepackage{upgreek}
\usepackage{makecell}
\usepackage{textcomp}
\usepackage[flushleft]{threeparttable}
\usepackage[symbol]{footmisc}

\usepackage{xcolor}
\usepackage{lineno}

\title{The Giant Magellan Telescope high contrast adaptive optics phasing testbed (p-HCAT): lab tests of segment/petal phasing with a pyramid wavefront sensor and a holographic dispersed fringe sensor (HDFS) in turbulence}

\author[a,b,*]{Alexander D. Hedglen}
\author[b]{Laird M. Close}
\author[a, $\dagger$]{Sebastiaan Y. Haffert}
\author[a]{Jared R. Males}
\author[a,b]{Maggie Kautz}
\author[c]{Antonin H. Bouchez}
\author[c]{Richard Demers}
\author[c]{Fernando Quir\'{o}s-Pacheco}
\author[c]{Breann N. Sitarski}
\author[a,b,d,e]{Olivier Guyon}
\author[a]{Kyle Van Gorkom}
\author[a]{Joseph D. Long}
\author[a,b]{Jennifer Lumbres}
\author[f]{Lauren Schatz}
\author[f]{Kelsey Miller}
\author[a,b]{Alex Rodack}
\author[a,b]{Justin M. Knight}

\affil[a]{University of Arizona, Steward Observatory, Tucson, AZ, USA}
\affil[b]{University of Arizona, College of Optical Sciences, Tucson, AZ, USA}
\affil[c]{GMTO Corp., Pasadena, CA, USA}
\affil[d]{Subaru Telescope, National Astronomical Observatory of Japan}
\affil[e]{Astrobiology Center, National Institutes of Natural Sciences, Japan}
\affil[f]{Kirtland Air Force Base, Air Force Research Laboratory, Albuquerque, NM, USA}

\cftpagenumbersoff{figure}
\cftpagenumbersoff{table} 
\begin{document} 
\maketitle

\begin{abstract}
The Giant Magellan Telescope (GMT) design consists of seven circular 8.4-m diameter mirrors, together forming a single 25.4-m diameter primary mirror. This large aperture and collecting area can help extreme adaptive optics (ExAO) systems such as GMT's GMagAO-X achieve the small angular resolutions and contrasts required to image habitable zone earth-like planets around late type stars and possibly lead to the discovery of life outside of our solar system. However, the GMT primary mirror segments are separated by large $>$ 30\,cm gaps, creating the possibility of fluctuations in optical path differences (piston) due to flexure, segment vibrations, wind buffeting, temperature effects, and atmospheric seeing. In order to utilize the full diffraction-limited aperture of the GMT for high-contrast natural guide star adaptive optics (NGSAO) science, the seven mirror segments must be co-phased to well within a fraction of a wavelength. The current design of the GMT involves seven adaptive secondary mirrors, a slow ($\sim$\,0.03\,Hz) off-axis dispersed fringe sensor (part of the Acquisition Guiding and Wavefront Sensing System's [AGWS] active optics off-axis guider), and a pyramid wavefront sensor (PyWFS; part of the Natural Guide star Wavefront Sensor [NGWS] adaptive optics) to measure and correct the total path length between segment pairs, but these methods have yet to be tested “end-to-end” in a lab environment. We present the design and working prototype of a “GMT High Contrast adaptive Optics phasing Testbed” (p-HCAT) which leverages the existing MagAO-X ExAO instrument to demonstrate segment phase sensing and simultaneous AO-control for high-contrast GMT natural guide star science (i.e., testing the NGWS wavefront sensor [WFS] architecture). We present the first test results of closed-loop piston control with one GMT segment using MagAO-X's PyWFS with and without simulated atmospheric turbulence. We show that the PyWFS was able to successfully control segment piston without turbulence within 12-33\,nm\,RMS for 0\,$\uplambda$/D\,--\,5\,$\uplambda$/D modulation, but was unsuccessful at controlling segment piston with generated $\sim$\,0.6 arcsec (median seeing conditions at the GMT site) and $\sim$\,1.2 arcsec seeing turbulence due to non-linear modal cross-talk and poor pixel sampling of the segment gaps on the PyWFS detector. These results suggest that a PyWFS alone is not an ideal piston sensor for the GMT (and likely the TMT and ELT). Hence, a dedicated ``second channel" piston sensor is required. We report the success of an alternate solution to control piston using a novel Holographic Dispersed Fringe Sensor (HDFS) while controlling all other modes with the PyWFS purely as a slope sensor (piston mode removed). This ``second channel" WFS method worked well to control segment piston to within 50\,nm\,RMS and $\pm$\,10\,$\upmu$m dynamic range under simulated 0.6 arcsec atmospheric seeing (median seeing conditions at the GMT site). These results led to the inclusion of the HDFS as the official second channel piston sensor for the GMT NGWS WFS. This HDFS + PyWFS architecture should also work well to control piston petal modes on the ELT and TMT telescopes.
\end{abstract}


{\noindent \footnotesize\textbf{*}Alexander D. Hedglen,  \linkable{ahedglen@optics.arizona.edu}
\newline
\noindent \footnotesize\textbf{$\dagger$} NASA Hubble Fellow}

\begin{spacing}{2}   

\section{Introduction}
\label{sect:intro}  
Our galaxy hosts $\sim$ 300 billion stars, each likely to have at least one exoplanet orbiting around it \cite{bryson_2021}. The release of Kepler data and other missions have led to $\sim$\,4,500 confirmed exoplanets,\footnote[1]{see \url{https://exoplanetarchive.ipac.caltech.edu/}} several of which fall into the potentially habitable, terrestrial size category \cite{thompson_2018}. If we want to answer one of mankind’s biggest questions, “Are we alone?” we need to detect signs of life (spectral biomarkers) by directly imaging these potentially habitable exoplanets in reflected light. This is a difficult task which requires Extremely Large Telescopes (ELTs; $\sim$ 30\,m in diameter) to achieve high angular resolutions and contrasts, extreme adaptive optics (ExAO) to suppress the effects of atmospheric seeing, and coronagraphy to block the starlight \cite{guyon_exao}. These three technologies could coexist once the 25.4-m Giant Magellan Telescope (GMT) is completed in 2029 \cite{males_gmagaox}. With the combined power of ExAO and the future GMT, the discovery of life outside of our solar system may become a reality. However, the GMT’s unique seven segmented M1 and M2 design raises a challenge to keep the telescope segments co-phased---a task which is critical for direct imaging of exo-earths and any other diffraction-limited science with the GMT.

The GMT consists of seven 8.4-m primary mirror segments (see Figure \ref{fig:figure1}) that are separated by $>$ 30\,cm gaps, inevitably creating co-phasing errors between the segments that limit the telescope's performance. Temperature changes, gravity load, segment vibrations, wind buffeting \cite{pacheco}, and atmospheric seeing \cite{schwartz_2017} will cause the optical path difference (OPD) between these segments to fluctuate as much as tens of microns, but for high-contrast Natural Guide Star Adaptive Optics (NGSAO) science, the segments must be co-phased to well within a fraction of a wavelength to achieve diffraction-limited performance and allow an ExAO system's coronagraph to properly block out the light from a star and image an exoplanet (see Section \ref{sect:risks}).

The initial phasing strategy of the GMT for NGSAO science involved the AGWS off-axis dispersed fringe sensor (DFS) and two different ``color" pyramid wavefront sensors (PyWFSs) to measure and control the OPD between segment pairs \cite{pacheco}. The two PyWFSs were intended to operate at different wavelengths (a broad band channel and a narrow band ``second channel") to overcome 2$\uppi$ phase ambiguities and increase the capture range of the WFS \cite{pinna_2006}. In NGSAO mode, the off-axis slow ``active optics" DFS would be used to reduce the initial OPD to within capture range of the two PyWFSs, which could theoretically control segment piston errors to within 30\,nm\,RMS while also correcting for atmospheric turbulence \cite{pinna_2014}. This architecture needed to be tested with real optics, AO hardware, and sensors in a lab environment to demonstrate that this NGWS phasing strategy would prove successful for the GMT.

We present the design and working prototype of a GMT High Contrast adaptive Optics phasing Testbed (HCAT) that simulates piston co-phasing errors between the GMT segments with real optics and leverages the existing AO instrument, MagAO-X \cite{males_2020, close_2018}, to test a PyWFS's ability to control piston co-phasing errors between the GMT segments while simultaneously correcting for atmospheric turbulence. We introduce a novel piston sensor optic that we named the ``Holographic Dispersed Fringe Sensor (HDFS; see Haffert et al.) \cite{haffert_2022}, which is included in the HCAT design as a ``second channel" for the testbed to sense piston and help resolve 2$\uppi$ phase ambiguities. Here we show the design and build of the prototype testbed---p-HCAT (completed in 2021)---and the first results of closed-loop piston control with one out of four GMT segments using MagAO-X's PyWFS and the novel HDFS (see Section \ref{sect:results}). These results from p-HCAT proved that the HDFS is a powerful piston sensor in closed-loop with the PyWFS working purely as a slope sensor. This new WFS method worked well to control piston to within 50\,nm\,RMS and $\pm$\,10\,$\upmu$m dynamic range under simulated 0.6 arcsec atmospheric seeing (median seeing conditions at the GMT site \cite{thomas-osip-2008}). Hence, this new HDFS + PyWFS piston sensing architecture is now the official phasing strategy for the GMT NGWS WFS. The dual PyWFS technique mentioned earlier is no longer being pursued by the GMT.

Section \ref{sect:project-goals} gives an overview of the project goals and risks. Section \ref{sect:previous-work} discusses other phasing testbeds that exist. A review of the MagAO-X instrument is given in section \ref{sect:magao-x}. Section \ref{sect:p-hcat} describes the optical/optomechanical design of p-HCAT and introduces the HDFS. Section \ref{sect:results} reports the results of closed-loop piston control with the PyWFS and HDFS. Section \ref{sect:HCAT} explores the design and future plans for the full-scale ``stage 2" testbed that will simulate all seven GMT segments with six off-axis piston, tip, and tilt segments and briefly mentions the current plans for the last stage of the HCAT project: feeding HCAT into the GMT's Natural Guide Star Wavefront Sensor prototype (NGWS-P), currently being developed by INAF - Arcetri in Italy. 

\section{Project Goals and Risks}
\label{sect:project-goals}

\subsection{HCAT Goals}
\begin{figure}[h!]
    \centering
    \includegraphics[width = 0.95\textwidth, angle = 0]{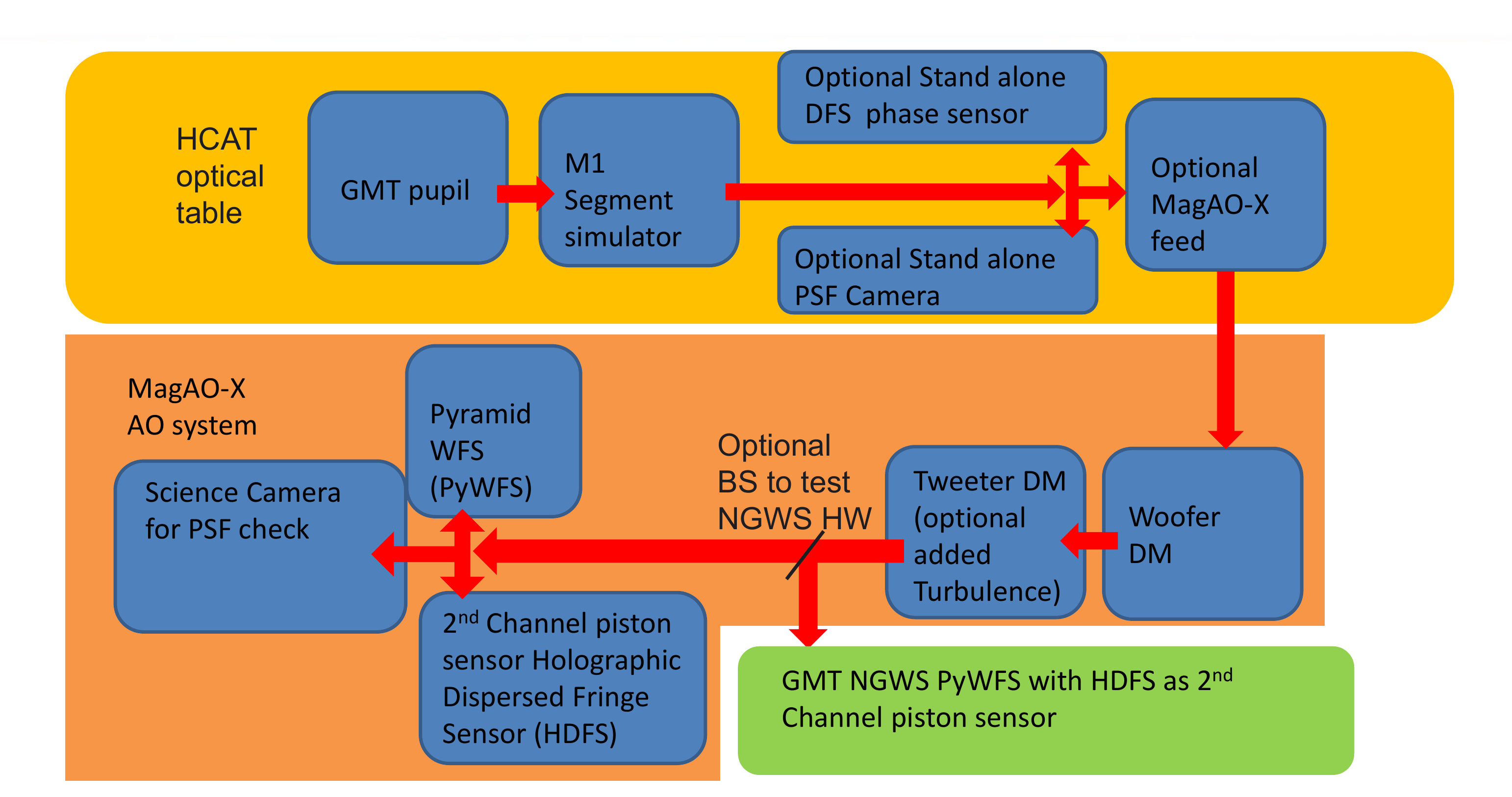}
    \\[6pt]
    \caption{A simple flowchart summarizing the HCAT (and p-HCAT) testbed project. The testbed simulates a GMT pupil with one out of four variable piston, tip, and tilt segments for stage 1 (p-HCAT) and eventually six out of seven piston, tip, and tilt segments for stage 2 (HCAT). The testbed is designed to feed light into the existing ExAO instrument MagAO-X for piston AO control with a PyWFS and a second channel HDFS piston sensor, but we also have a standalone operational mode for phase sensing without MagAO-X if needed. For stage 3 in 2023 we will feed light into the prototype NGWS to test the GMT's internal AO hardware and software for NGSAO wavefront sensing.}
    \label{fig:flowchart}
\end{figure}

The overall goal of the HCAT testbed is to retire the GMT high-risk item of \emph{phasing performance}. HCAT will simulate real phasing errors between the GMT segments in a lab environment and test an ExAO system's ability to control segment piston errors in the presence of atmospheric turbulence. See Figure \ref{fig:flowchart} for a simple flowchart that summarizes the HCAT testbed project. 

The full scope of the HCAT project consists of three stages:
\begin{enumerate}
    \item \textbf{Stage 1} (\emph{completed in 2021}): Build a simple preliminary testbed (p-HCAT) that simulates one out of four GMT segments that can piston, tip, and tilt and use the existing MagAO-X AO system to test phase sensing of one GMT segment with a PyWFS and a second channel prototype HDFS piston sensor under realistic atmospheric turbulence conditions.  
    \item \textbf{Stage 2} (\emph{2022}): Build the full-scale testbed (HCAT) which includes all seven GMT segments with six that can piston, tip, and tilt and test a working concept for the GMagAO-X ``parallel DM" \cite{close_gmagaox}. Perform phase sensing demonstrations with MagAO-X and the new PyWFS + HDFS architecture.
    \item \textbf{Stage 3} (\emph{2023}): Use HCAT and MagAO-X to feed the GMT's Natural Guide Star Wavefront Sensor prototype (NGWS-P), test its internal PyWFS + HDFS architecture, and implement control algorithms provided by GMTO.
\end{enumerate}

\subsection{Risks}
\label{sect:risks}
The U.S. Decadal Survey on Astronomy and Astrophysics 2020 (Astro2020) agrees that the highest technical risk for the GMT is the phasing and alignment of the primary mirror segments \cite{astro2020}. Phasing errors in the form of piston are of particular concern, since traditional slope wavefront sensors like the Shack-Hartmann cannot sense piston across discontinuities in the telescope pupil. The atmosphere alone will introduce differential segment piston (or ``petal") errors between GMT segments that can be on the order of $\pm$\,5\,$\upmu$m \cite{schwartz_2017, bertrou_2021}. Furthermore, the large gaps between the GMT segments span multiple atmospheric coherence lengths $r_0$ (for $\uplambda <$\,1.2\,$\upmu$m), so atmospheric turbulence will have a differential phase error across these gaps that can be as much as 1\,$\uplambda$ in median seeing conditions \cite{schwartz_2017}. Visible to near-infrared AO systems (like GMagAO-X \cite{males_gmagaox, close_gmagaox}) will have wavefront sensor pixels in these gaps, causing a discontinuity in the fitted wavefront data that will generate differential piston between the GMT segments (or ``petals") \cite{bertrou_2021}. This is a problem not only for the GMT, but for all Giant Segmented Mirror Telescopes (GSMTs) as well. 

\begin{figure}[h!]
    \centering
    \includegraphics[width = 0.99\textwidth, angle = 0]{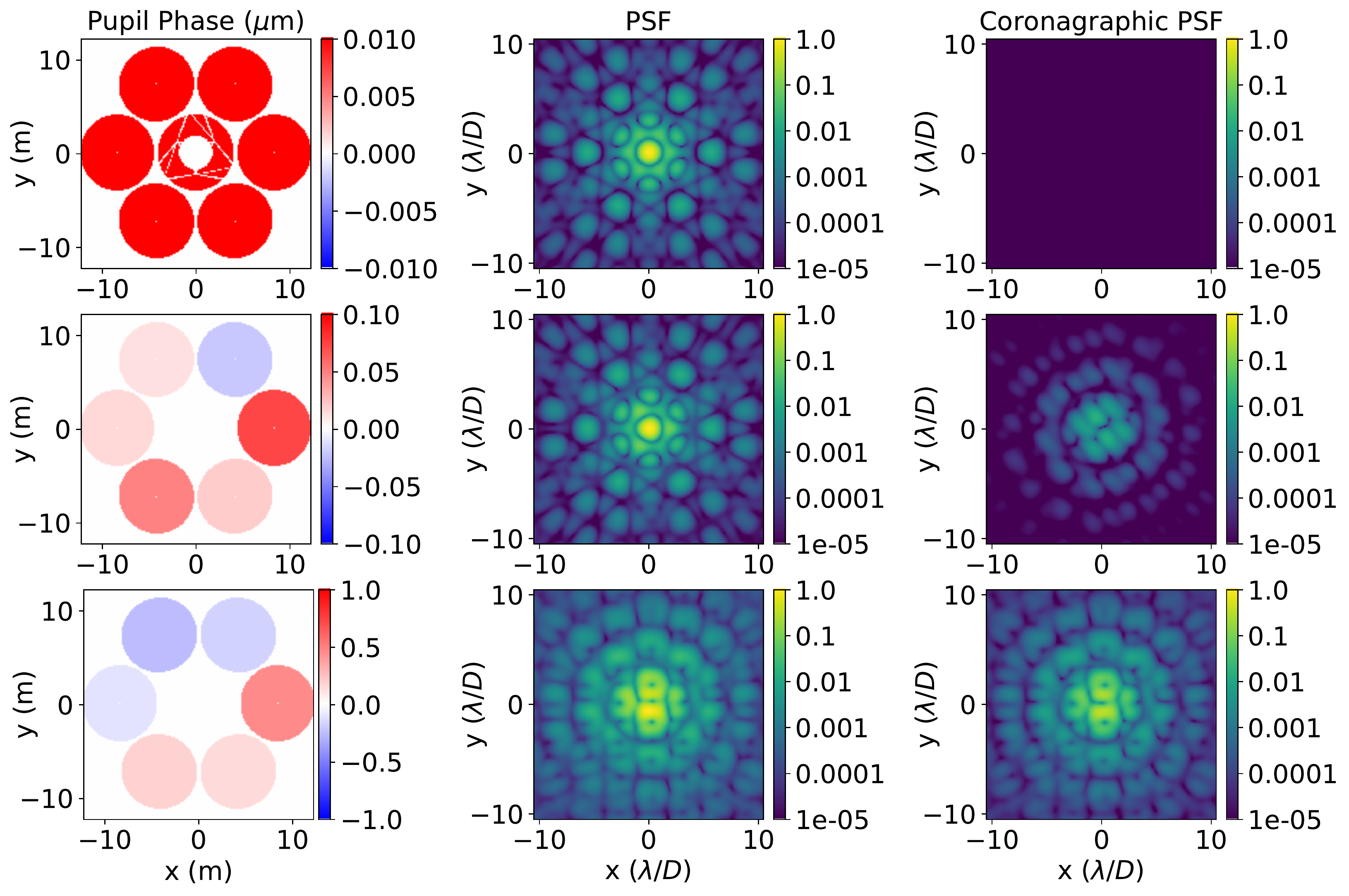}
    \\[6pt]
    \caption{A simulation showing the comparison of different levels of piston errors between the GMT segments and the resulting PSF and coronagraphic PSF (assuming a perfect coronagraph and no atmospheric turbulence) at $\uplambda_c$ = 800\,nm, 100\,nm bandwidth. (Top row): the GMT segments have zero differential piston error, so the perfect coronagraph is able to block out the light from the star perfectly. (Middle row): the GMT segments have 30\,nm RMS piston error. The PSF has a Strehl ratio of 95\% and the coronagraph is still functional but light is starting to leak out of the coronagraph (see Figure \ref{fig:coronagraph-plots} for more details). (Bottom row): the GMT segments have 300\,nm RMS piston error. The PSF  has a Strehl ratio of 0\% and most of the light is leaking out of the coronagraph, making it impossible to carry out any diffraction-limited science cases for the GMT.}
    \label{fig:figure1}
\end{figure}

\begin{figure}[h!]
    \centering
    \begin{tabular}{cc}
    \includegraphics[page = 1, width = 0.47\textwidth, angle = 0]{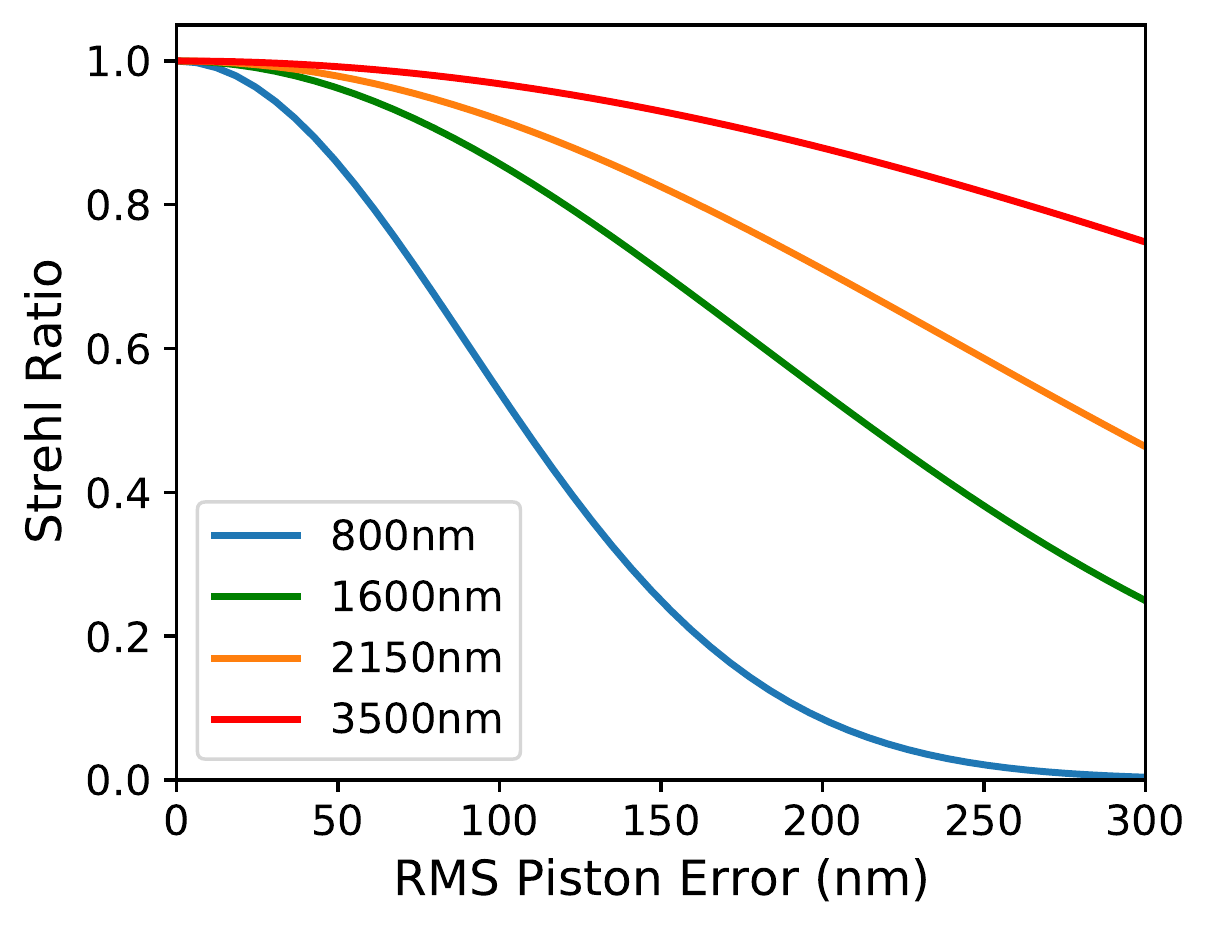} &
    \includegraphics[page = 2, width = 0.47\textwidth, angle = 0]{Figure3.pdf} \\
    \hspace{0.9cm} (a) & \hspace{0.9cm} (b)\\
    \end{tabular}
    \\[6pt]
    \caption{(a) The Strehl ratio of the PSF from Figure \ref{fig:figure1} as a fuction of RMS piston error for a few different wavelengths. In order to achieve high Strehl in the visible, the GMT segments must be co-phased to well within a fraction of a wavelength. (b) The radial profiles of the coronagraphic PSF from Figure \ref{fig:figure1} for different levels of piston errors (no other errors). We can see that in order to achieve high-contrast, the segments must be co-phased as well as possible.}
    \label{fig:coronagraph-plots}
\end{figure}

Figure\,\ref{fig:figure1} shows an example of how differential piston errors between the GMT segments could affect the PSF and coronagraphic PSF, assuming a perfect coronagraph \cite{cavarroc_2006} and no atmospheric turbulence for $\uplambda_\text{c}$\,=\,800\,nm, 100\,nm bandwidth. Figure \ref{fig:coronagraph-plots} shows the Strehl ratio of the PSF for a few different wavelengths and the radial profiles of the coronagraphic PSFs for the same conditions. The coronagraphic PSFs were all normalized by the reference PSF to show the relative contrast.
These simulations were performed with High Contrast Imaging for Python (HCIPy), an open source AO and coronagraph simulator \cite{por_2018}. From these simulations, we can see why the GMT segments must be co-phased to well within a fraction of a wavelength to achieve high Strehl ratio and high contrast in the visible to near-infrared wavelengths.

\section{Other ELT Phasing Testbeds}
\label{sect:previous-work}
The Miniscule ELT (MELT) is a phasing testbed designed by the European Southern Observatory for developing wavefront control and phasing strategies for the European ELT (E-ELT) \cite{melt_2018}. MELT was derived from the Active Phasing Experiment (APE), which tested several different optical phasing sensors on-sky with the Very Large Telescope \cite{ape_2008}. These phase sensors were designed with the intention of calibrating hexagonally segmented ELTs on timescales of $\sim$\,24\,hours while the edge sensors keep the segments co-phased in real time \cite{vigan_2011}. This is similar to the W. M. Keck Observatory's technique of keeping its primary mirror segments phased \cite{chanan_1998, chanan_2000}. 

The GMT will have $\sim$\,100 times larger gaps between M1 segments than the E-ELT, and the GMT M1 segments are also made of borosilicate glass instead of zerodur, so its edge sensors are not expected to be stable for more than a few minutes \cite{bouchez_2012}. For this reason, the GMT's off-axis AGWS was designed to maintain segment alignment (but not correct atmospheric piston) using a slow ($\sim$\,0.03\,Hz) off-axis DFS \cite{van-dam-2016}. This slow, seeing-limited DFS is sufficient for ground layer AO and laser tomography AO science modes, but it is not sufficient for NGSAO mode. A prototype of the DFS was tested on-sky at the Magellan-Clay telescope in 2016\cite{kopon-2016} and a full-scale laboratory testbed for the AGWS (the GMT Wide Field Phasing Testbed) is currently being developed at the Smithsonian Astrophysical Observatory.

\section{MagAO-X}
\label{sect:magao-x}
\begin{figure}[b!]
    \centering
    \includegraphics[width = 1.0\textwidth, angle = 0]{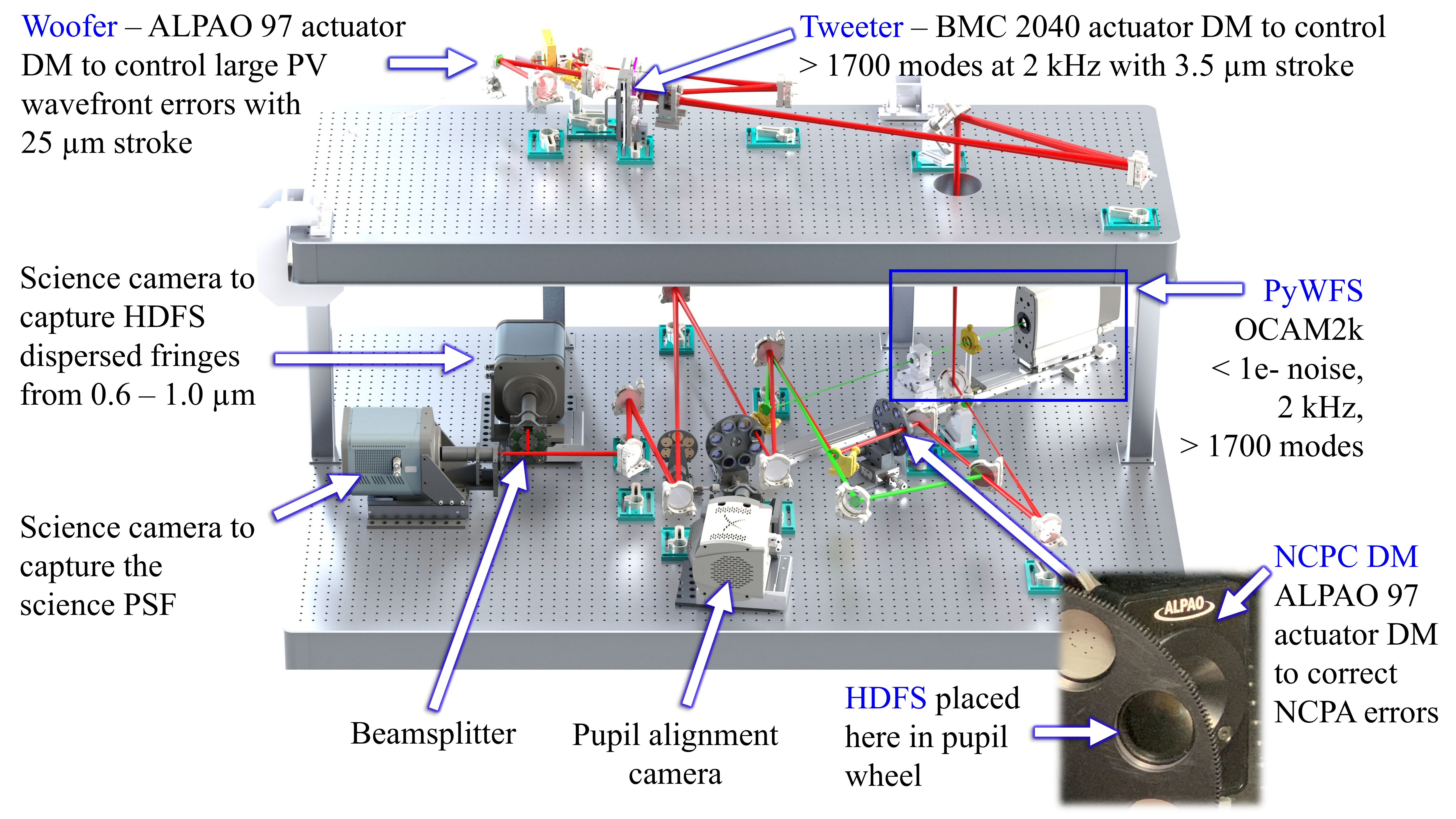}
    \\[6pt]
    \caption{The MagAO-X instrument. MagAO-X uses a PyWFS and three DMs (a 2,040 actuator ``tweeter" DM, 97 actuator ``woofer" DM, and 97 actuator ``non-common path corrector" (NCPC) DM to correct for atmospheric turbulence and non-common path aberrations (NCPA) at visible wavelengths.}
    \label{fig:magao-x}
\end{figure}

The Magellan Extreme Adaptive Optics system (MagAO-X) is a new visible-to-near-IR ExAO instrument designed for the 6.5-m Magellan Clay telescope in Chile that was recently built (first light in fall 2019) in the University of Arizona's Center for Astronomical Adaptive Optics (CAAO) Extreme Wavefront Control Lab \cite{males_2020, males_2018, close_2018}. Figure \ref{fig:magao-x} shows a detailed view of the MagAO-X instrument. MagAO-X utilizes a PyWFS and three deformable mirrors (DMs) to produce images with high Strehl ratio ($>$ 70\%), high resolution (19\,mas), and high planet/star contrasts ($<$ 10$^{-4}$) at 656\,nm to measure and correct atmospheric distortions and image exoplanets with high precision \cite{close_2020}. 
The three DMs in MagAO-X consist of a Boston Micromachines Corp. (BMC) 2,040 actuator ``tweeter" DM to control $>$ 1700 modes at 2\,kHz speeds with 3.5\,$\upmu$m stroke, an ALPAO 97 actuator ``woofer" DM to control large peak-to-valley (PV) low-order wavefront errors with 25\,$\upmu$m stroke, and another ALPAO 97 actuator ``non-common path corrector" (NCPC) DM to control non-common path aberrations (NCPA). These three DMs were optimized via a focus diversity phase retrieval method to bring the internal static wavefront error of MagAO-X down to a total of 18.7\,nm RMS \cite{van-gorkom}. HCAT is designed to feed MagAO-X (while it is in the lab between observing runs in Chile) so that we can demonstrate ExAO control with a ``GMT-like" pupil. We incorporate the novel HDFS by simply placing the optic in one of MagAO-X's pupil wheels (see Figure \ref{fig:magao-x}).

\section{p-HCAT}
\label{sect:p-hcat}

The proto-High Contrast daptive Optics phasing Testbed (p-HCAT) was developed as the first stage of the HCAT project to create a simple four segment M1 GMT simulator that only has one variable piston, tip, and tilt segment. The fabrication and alignment of p-HCAT was completed in spring 2021 (just 2 months into the formal project timeline). This testbed allowed us to rapidly test the PyWFS as a phasing sensor and was critical to inform the GMT of the HDFS as a possible choice for a second channel phasing sensor. P-HCAT has also helped pave the way for the full HCAT build (explained in Section \ref{sect:HCAT}), which will include all seven M1 GMT segments with six variable piston, tip, and tilt elements. In this section we discuss the optical design of p-HCAT and introduce the HDFS.

\subsection{Optical Design}
\label{sect:optical-design}
\begin{figure}[t!]
    \centering
    \includegraphics[width = 1.0\textwidth, angle = 0]{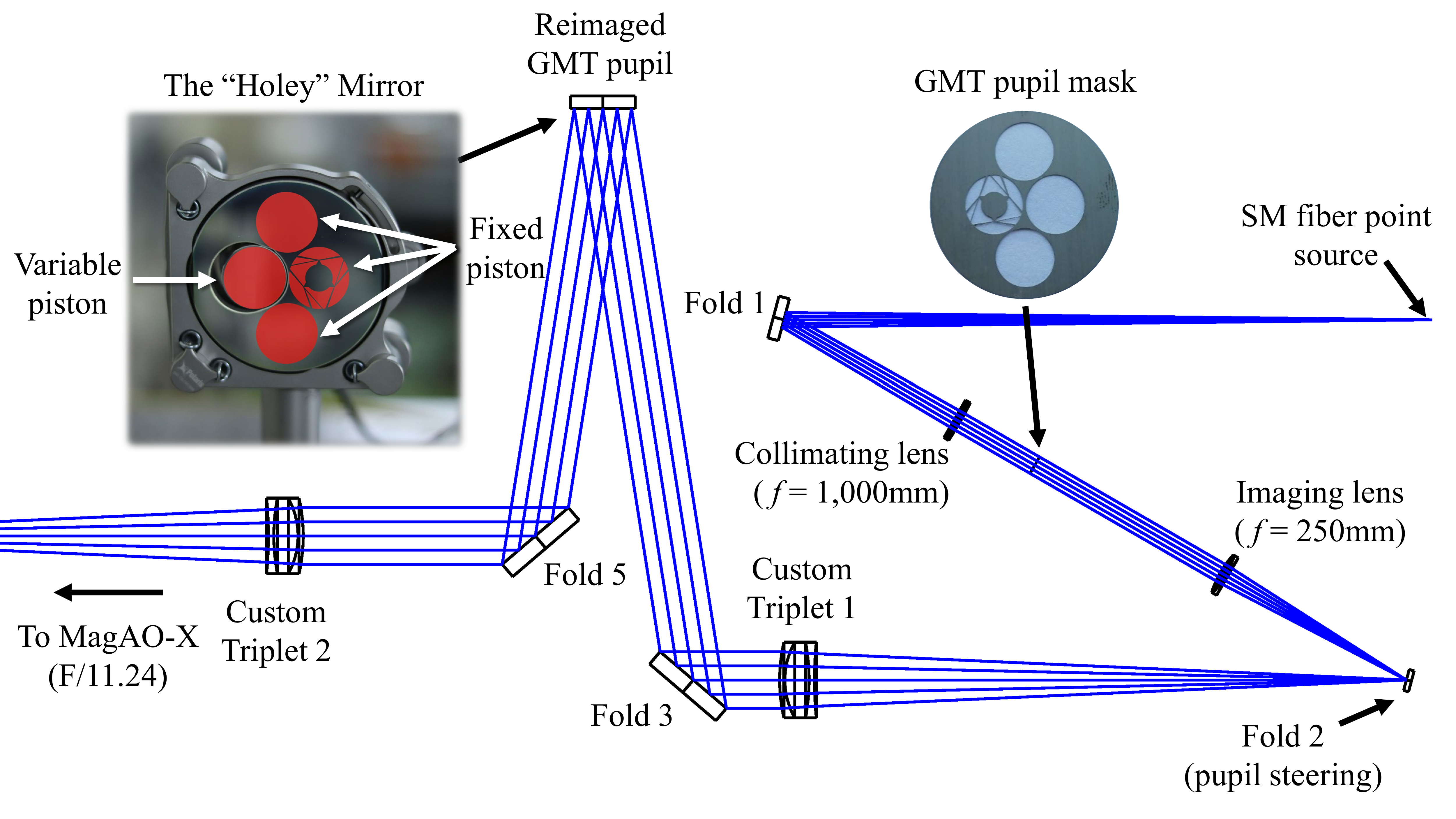}
    \\[6pt]
    \caption{The optical design of p-HCAT. A four-segment GMT pupil mask is re-imaged onto the surface of the ``holey" mirror where piston errors are introduced to one of the four segments with a PI S-325 piezo (tip/tilt and piston) actuator.}
    \label{fig:figure2}
\end{figure}

\begin{figure}[h!]
    \centering
    \includegraphics[width = 1.0\textwidth, angle = 0]{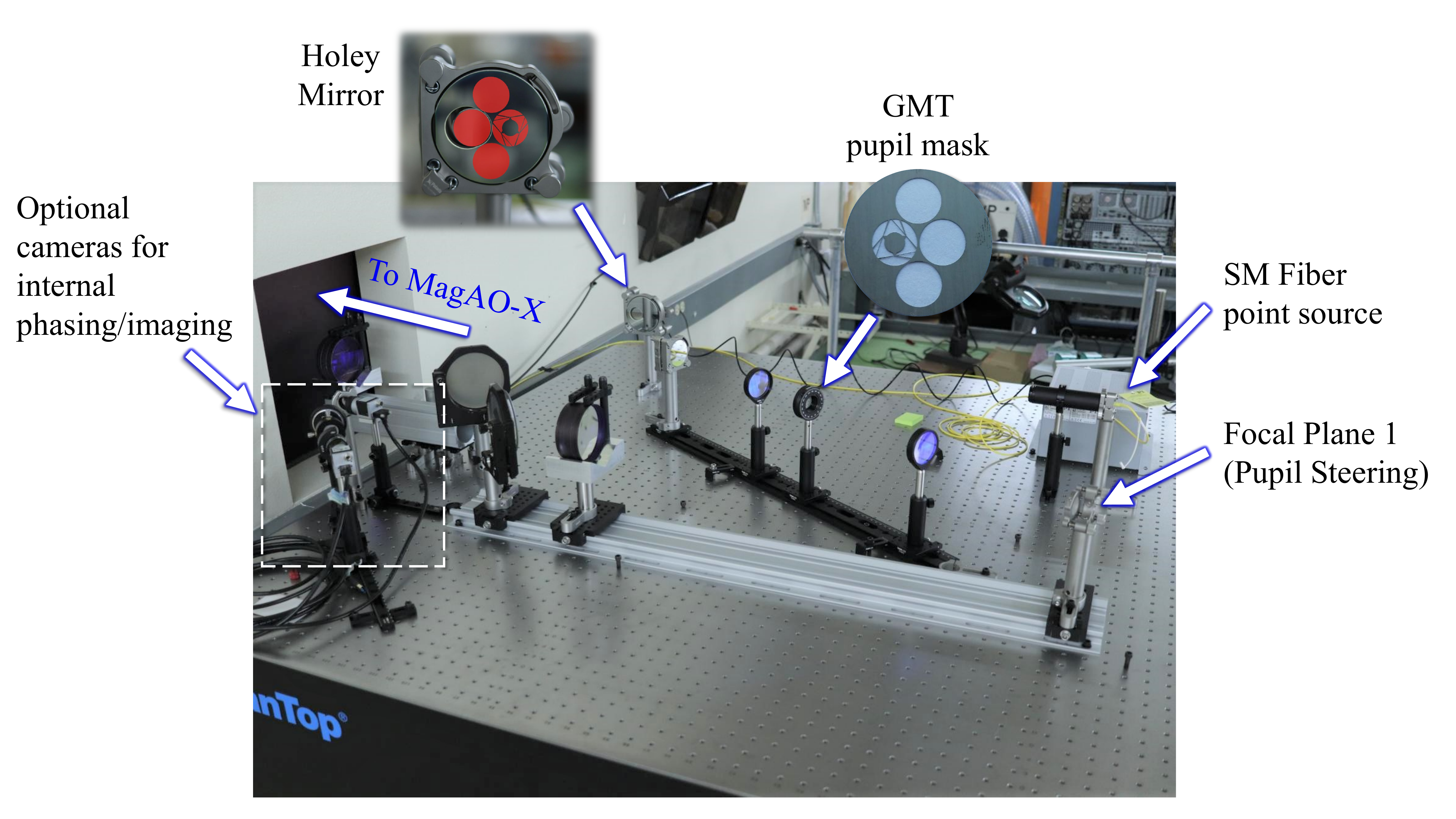}
    \\[6pt]
    \caption{p-HCAT as built in the lab. Light exits through a hole in the wall and into MagAO-X in the adjacent lab. Optionally, a fold mirror may be placed after custom triplet \#2 for internal testing when needed.}
    \label{fig:figure3}
\end{figure}

\begin{figure}[h!]
    \centering
    \includegraphics[width = 0.9\textwidth, angle = 0]{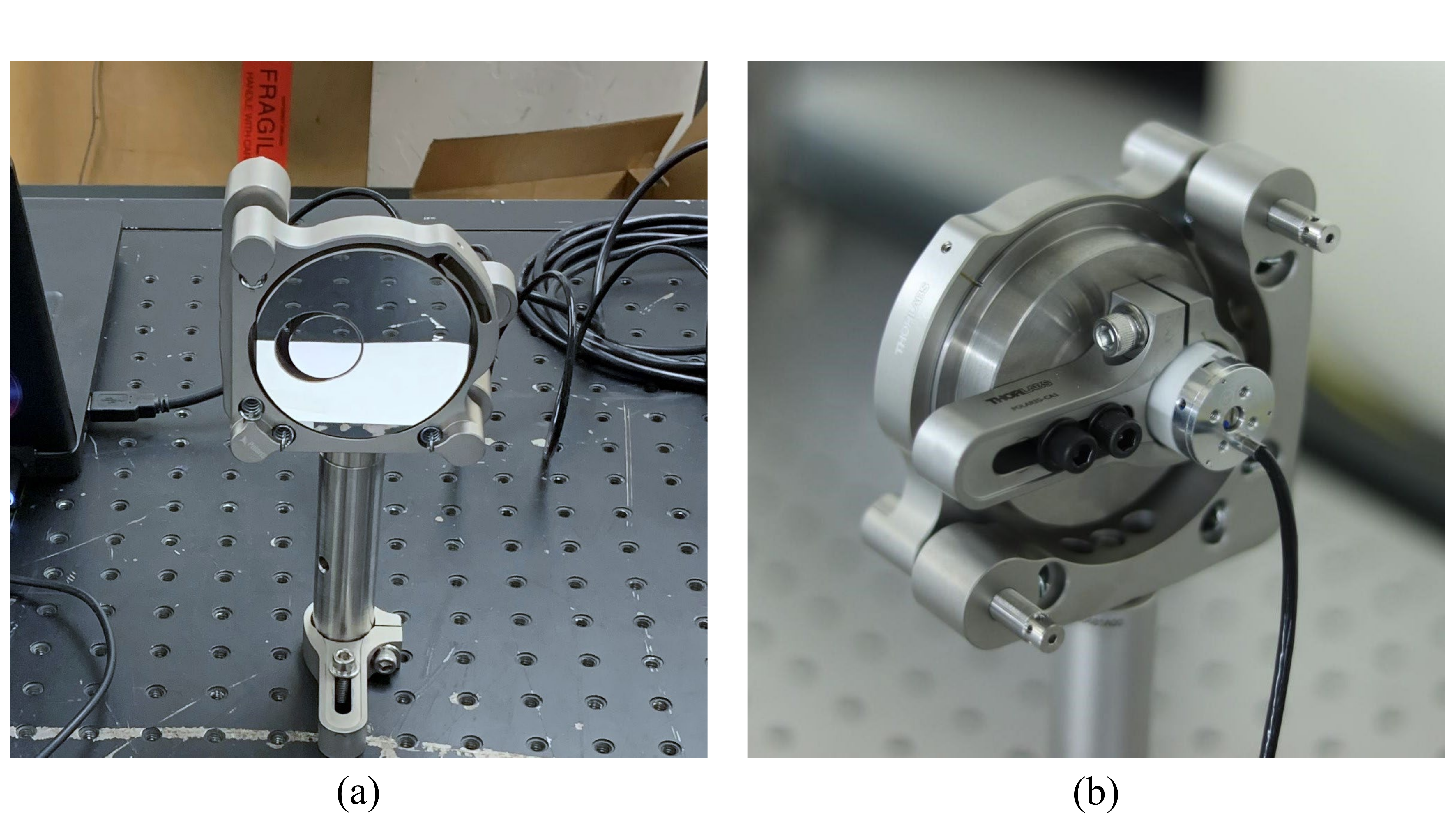}
    \\[6pt]
    \caption{(a) The Holey Mirror front view with the 1.0-inch mirror segment co-phased. (b) The Holey Mirror rear view. A 1.0-inch mirror is glued to a PI S-325 actuator which protrudes through a hole in the 3.0-inch mirror exactly over one of the re-imaged GMT segments. The piezo actuator has 30\,$\upmu$m mechanical piston range and 4\,mrad mechanical tip/tilt range.}
    \label{fig:holey-mirror}
\end{figure}

The optical design of p-HCAT is shown in Figure \ref{fig:figure2} while Figure \ref{fig:figure3} shows the testbed ``as built" in the lab. The design consists of two pupil relays: a commercial doublet lens relay and a custom triplet lens relay. In the first relay, a flat plane wave was simulated using a single-mode fiber point source and a $f$\,=\,1,000\,mm commercial doublet. A four-segment stainless steel etched GMT pupil mask was placed in the collimated beam to define the entrance pupil of the system while a \emph{f}\,=\,250\,mm commercial doublet lens creates the first focal plane image. The size of the pupil mask and the focal length of the imaging lens were chosen to create an F/\# of 11.24 as required for feeding MagAO-X. In this manner we effectively fill MagAO-X's input beam and we switch from the Magellan-Clay to the GMT. In the second pupil relay, a magnified image of the pupil mask is formed on the surface of a 3.0-inch mirror with a hole cut into it (i.e., ``the Holey Mirror," see Figure\,\ref{fig:holey-mirror}). This hole allows a 1.0-inch mirror glued to a piezo actuator to ``poke" through, exactly around one of the re-imaged GMT segments, so that one of the GMT segments can piston, tip, and tilt, while the other three segments are fixed in piston on the flat mirror. Light is then refocused to feed light at F/11.24 into MagAO-X.


  %

  %

\subsubsection{Custom Triplet Lenses}
\label{sect:custom-triplets}

\begin{figure}[b!]
    \centering
    \begin{tabular}{cc}
        \includegraphics[page = 1, width = 0.49\textwidth, angle = 0]{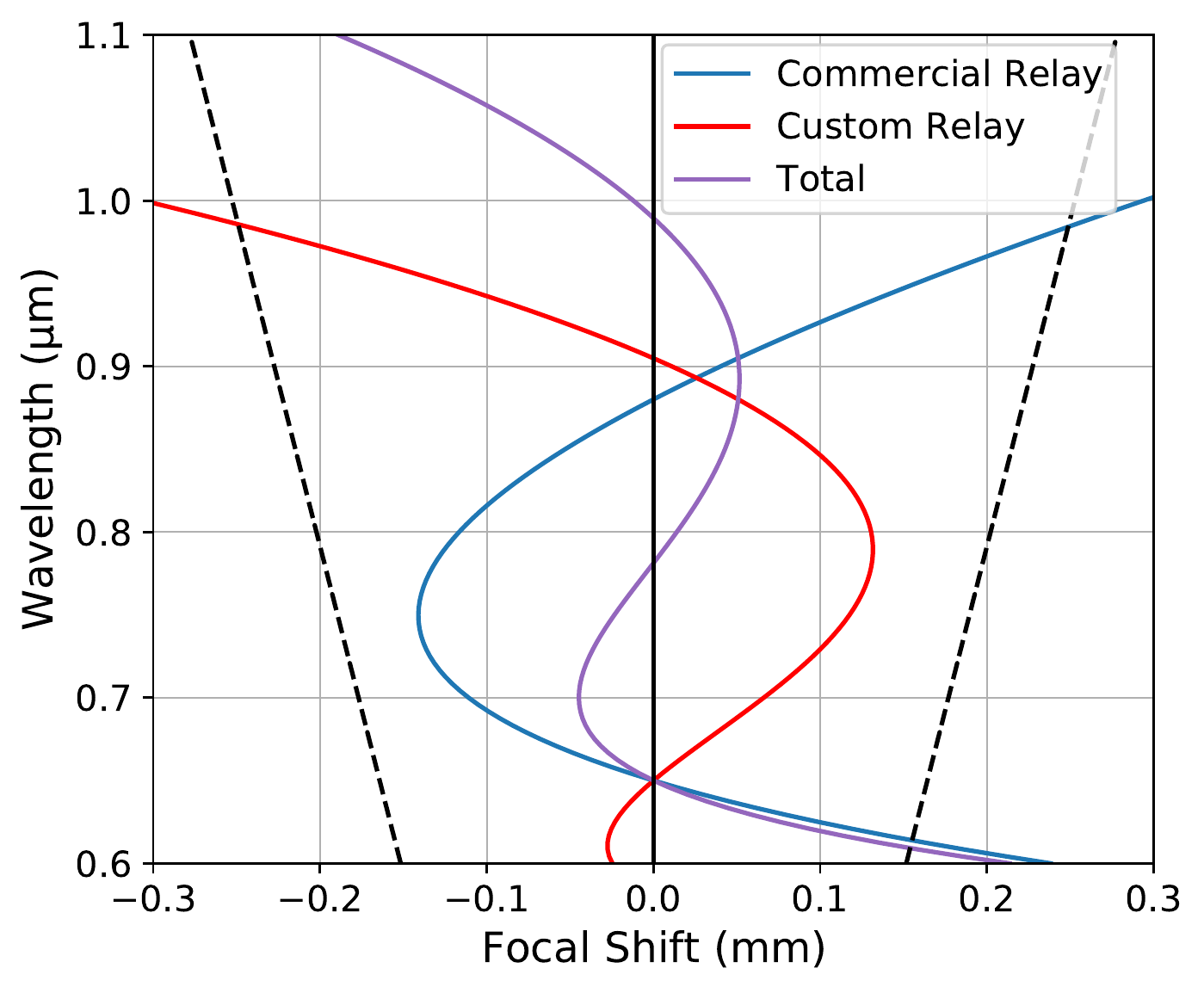} &
        \includegraphics[page = 2, width = 0.49\textwidth, angle = 0]{Figure8.pdf} \\ 
        \hspace{0.7cm} (a) & \hspace{0.7cm} (b) \\
    \end{tabular}
    \\[6pt]
    \caption{(a) Chromatic focal shift plots from the first pupil lens relay (blue) and the custom triplet lens relay (red). The total resulting focal shift is shown in purple. (b) The final focal shift plot for p-HCAT with the custom triplets optimized for minimizing wavefront error. The focal shift is still considered diffraction-limited from 0.6\,--\,1.1\,$\upmu$m. The diffraction-limited depth of focus is shown with the dashed lines and is given by $\pm$ $2 \lambda F^2$.}
    \label{fig:focal-shift}
\end{figure}

\begin{figure}[h!]
    \centering
    \includegraphics[width = 0.9\textwidth, angle = 0]{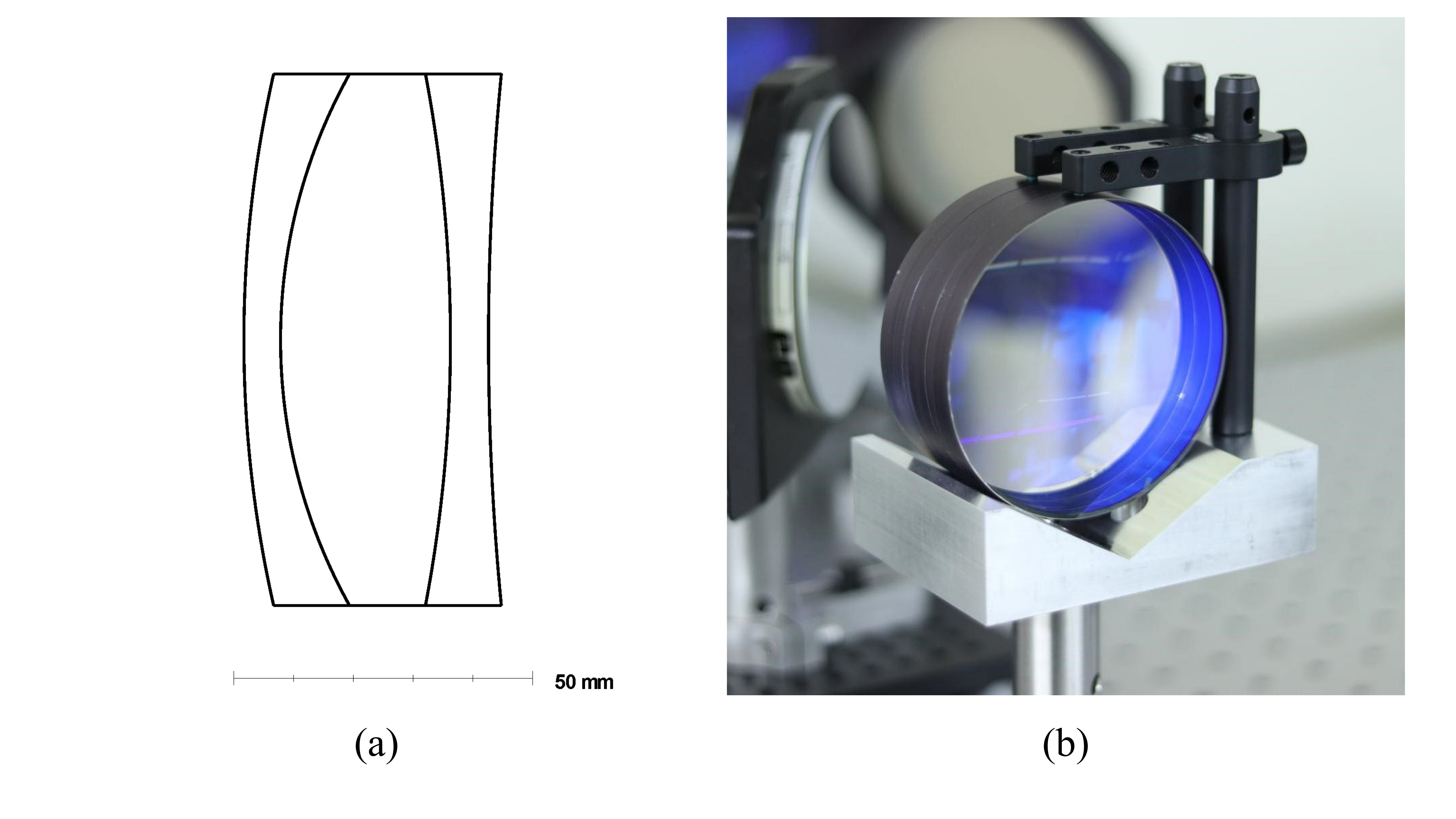}
    \\[6pt]
    \caption{(a) A cross-section of the custom triplet lens design. The orientation follows Table \ref{tab:lens-prescription} such that surfaces 1 -- 4 are from left to right. (b) The fabricated custom triplet lens mounted on the testbed. The triplet lens was designed to have an effective focal length of 750\,mm and an outer diameter of 3.50-inches.}
    \label{fig:custom-triplet}
\end{figure}

The MagAO-X instrument currently operates from 0.60\,$\upmu$m -- 1.10\,$\upmu$m, so p-HCAT was designed to create an achromatic beam within this wavelength range using a combination of commercial doublet lenses and custom triplet lenses. The custom triplet lenses were designed to counter the chromatic aberration from the first commercial doublet lens relay. The commercial doublet lens relay reduced the overall project cost, but required two custom triplet lenses for the second pupil relay to make the whole system achromatic. Figure\,\ref{fig:focal-shift}a shows the result of adding the chromatic focal shift plots from the first commercial lens relay with the new custom triplet lens relay.
Since the custom triplets were designed to have the opposite focal shift curve of the commercial lens relay, adding the focal shift curves together results in a final effective focal shift curve that is minimized (diffraction-limited from 0.60\,$\upmu$m\,--\,1.10\,$\upmu$m). Figure \ref{fig:focal-shift}b shows the final optimization of the triplet lens design (to minimize the total wavefront error rather than the chromatic focal shift). The focal shift changed slightly, however the final chromatic focal shift is still considered diffraction-limited from 0.60\,$\upmu$m\,--\,1.10 $\upmu$m. Each lens consists of the same design with an effective focal length of 750\,mm and an outer diameter of 3.50-inches. All surfaces are spherical with $<$\,$\uplambda$/10\,PV surface irregularity at 633\,nm. Table \ref{tab:lens-prescription} shows the custom triplet lens prescription while Figure\,\ref{fig:custom-triplet} shows a cross section of the lens design and an image of the finished lens on the testbed.

\begin{table}[ht]
        \caption{P-HCAT Custom Triplet Lens Prescription.} 
        \label{tab:lens-prescription}
        \begin{center}   
        \begin{tabular}{|c|c|c|c|} 
        \hline
        \rule[-1ex]{0pt}{3.5ex}  \textbf{Surface} & \textbf{Radius [mm]} & \textbf{Center Thickness [mm]} & \textbf{Material} \\
        \hline\hline
        \rule[-1ex]{0pt}{3.5ex}  1 & 200.839 & 6.136 & S-BSM81  \\
        \hline
        \rule[-1ex]{0pt}{3.5ex}  2 & 91.344 & 28.430 & S-BSM2   \\
        \hline
        \rule[-1ex]{0pt}{3.5ex}  3 & -238.128 & 6.367 & S-NBH5  \\
        \hline
        \rule[-1ex]{0pt}{3.5ex}  4 & 460.985 &  &   \\
        \hline
        \end{tabular}
        \end{center}
\end{table} 

\subsection{Standalone Mode}
\label{sect:standalone-mode}
\begin{figure}[b!]
    \centering
    \includegraphics[width = 1.0\textwidth, angle = 0]{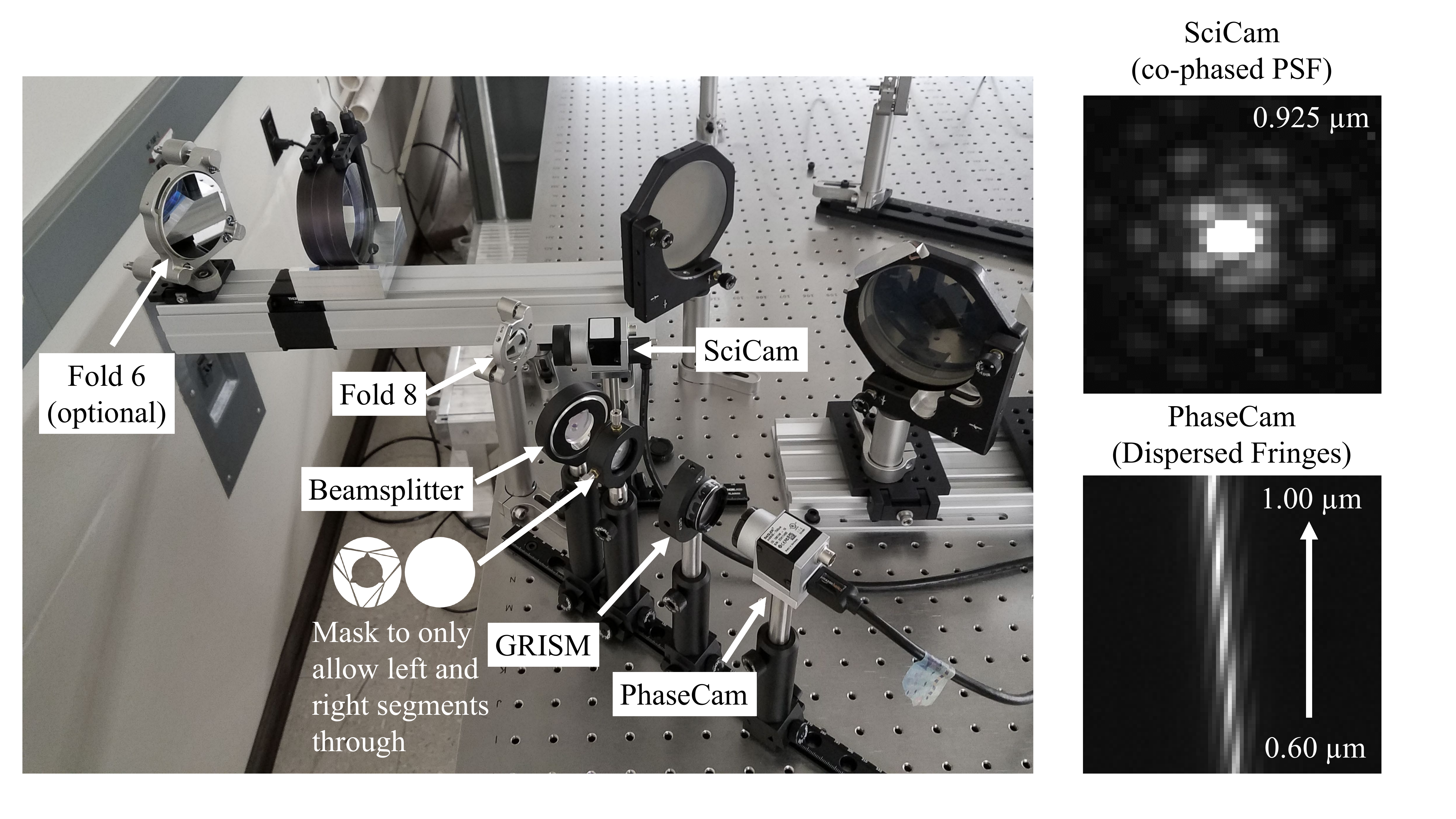}
    \\[6pt]
    \caption{p-HCAT standalone mode. A fold mirror may be placed after the last custom triplet lens to fold light back onto the testbed for internal testing without MagAO-X. Here a 50/50 beamsplitter splits the light into a science channel and a phasing channel to simultaneously observe the PSF and dispersed fringes for piston phasing of the GMT segment. The PSF is shown for zero piston error at 925\,nm while the phasing camera shows an example of dispersed fringes with several microns of piston error. With this much piston error, the dispersed fringes appear as a ``barber pole" since the phase wraps $\sim$\,5 times from the blue to the red. Their ability to linearly handle many 2$\uppi$ phase wraps makes dispersed fringe sensors powerful piston sensors with large dynamic ranges.}
    \label{fig:standalone-mode}
\end{figure}

To view the PSF in the final focal plane, an optional fold mirror may be placed after the final triplet lens to fold the light back onto the testbed (see Figure \ref{fig:standalone-mode}). This is our ``standalone mode" for initial alignment/phasing of the testbed without MagAO-X if needed. In standalone mode, a 50/50 beamsplitter splits the light into a science channel and a phasing channel. The phasing channel consists of a mask to block out the top and bottom GMT segments, only allowing the light from the piezo segment and the central obscuration segment through. 
A grism (prism + low resolution R\,$\sim$ 200\,grating) then disperses the light along a straight path (coma free) and dispersed fringes are created by the grism's first order in the focal plane for coarse piston phasing of the GMT segments.

\subsubsection{Error Budget and Strehl Ratio Measurements}
\label{sect:error-budget}

Each optical surface in p-HCAT was measured using a Zygo\textregistered{} interferometer to estimate the total static wavefront error, and from that, the Strehl ratio of the system in the final focal plane. Table\,\ref{tab:error} shows a summary of the measured RMS wavefront errors for each optical element and the total root sum squared (RSS) wavefront error. From these measurements we estimated the Strehl ratio at 925\,nm to be 78\%. The Holey Mirror dominates the total error due to slight warping of the mirror at the edge of the post-polished cored hole. 

\begin{figure}[ht!]
    \centering
    \includegraphics[width = 1.0\textwidth, angle = 0]{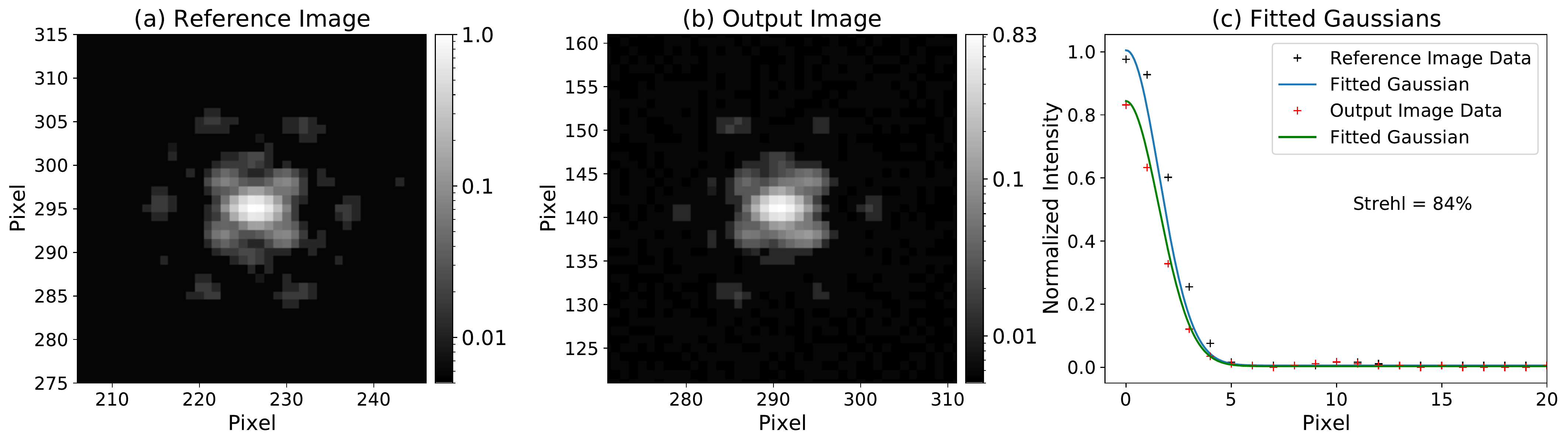}
    \\[6pt]
    \caption{(a) The reference PSF (first focal plane of p-HCAT) at $\uplambda_c$ = 925\,nm, 25\,nm bandwidth. (b) The output PSF (final focal plane of p-HCAT) at $\uplambda_c$ = 925\,nm, 25\,nm bandwidth. (c) The fitted radial profiles of the measured PSFs. The reference PSF is shown in blue while the output PSF is shown in green. The Strehl ratio was calculated to be 84\% from this data.}
    \label{fig:strehl}
\end{figure}

The Strehl ratio was then measured on the testbed, where we confirmed it fell within our error budget. The Strehl ratio was measured by comparing the PSF in the first focal plane (assumed to be 100\% Strehl at F/11.24) with the PSF in the final focal plane (also F/11.24). A temporary fold mirror was placed just before the pupil steering mirror to allow us to look at the PSF in the first focal plane. Since the same detector, wavelength, and F/\# were used at each focal plane, a ratio of the peaks normalized by the image flux is an accurate measurement of the delivered Strehl to MagAO-X. A non-linear least squares fit was used to fit a 2D Gaussian to the data to find the fitted peaks while the central 40 pixels were used to calculate the flux of the images. Figure\,\ref{fig:strehl} shows the radial profiles of the fitted PSF data. We measured a Strehl ratio of 84\% from this data, which matches the estimations from the error budget. From these results we concluded that the p-HCAT optical quality was sufficient enough to produce coherent images in the final focal plane as needed for performing phasing tests with MagAO-X.

\begin{table}[ht]
        \caption{p-HCAT Wavefront Error Budget.} 
        \label{tab:error}
        \begin{center}   
        \begin{tabular}{|c|c|c|} 
        \hline
        \rule[-1ex]{0pt}{3.5ex}  \textbf{Optical Element} & \textbf{\makecell{RMS Wavefront \\ Error [nm]}} & \textbf{Comments}  \\
        \hline\hline
        \rule[-1ex]{0pt}{3.5ex}  Custom Triplet 1 & 11 & \makecell{as measured by Optimax}  \\
        \hline
        \rule[-1ex]{0pt}{3.5ex}  Fold Mirror 3 & 22.8 & \makecell{as measured on Zygo}   \\
        \hline
        \rule[-1ex]{0pt}{3.5ex}  Holey Mirror & 59.5 & \makecell{as measured on Zygo}   \\
        \hline
        \rule[-1ex]{0pt}{3.5ex}  Fold Mirror 5 & 18.9 & \makecell{as measured on Zygo}  \\
        \hline
        \rule[-1ex]{0pt}{3.5ex}  Custom Triplet 2 & 11 & \makecell{as measured by Optimax}  \\
        \hline
        \rule[-1ex]{0pt}{3.5ex}  Fold Mirror 6 & 18.9 & \makecell{as measured on Zygo}  \\
        \hline
        \rule[-1ex]{0pt}{3.5ex}  Fold Mirror 7 & 10.1 & \makecell{as measured on Zygo}  \\
        \hline
        \rule[-1ex]{0pt}{3.5ex}  Fold Mirror 8 & 14.1 & \makecell{as measured on Zygo}  \\
        \hline
        \rule[-1ex]{0pt}{3.5ex}  \textbf{Total} & \textbf{72.9} & root sum squared\\
        \hline
        \end{tabular}
        \end{center}
\end{table} 

  %

\subsection{Feeding MagAO-X}
\label{sect:feeding-magao-x}

\begin{figure}[ht!]
    \centering
    \includegraphics[width = 1.0\textwidth, angle = 0]{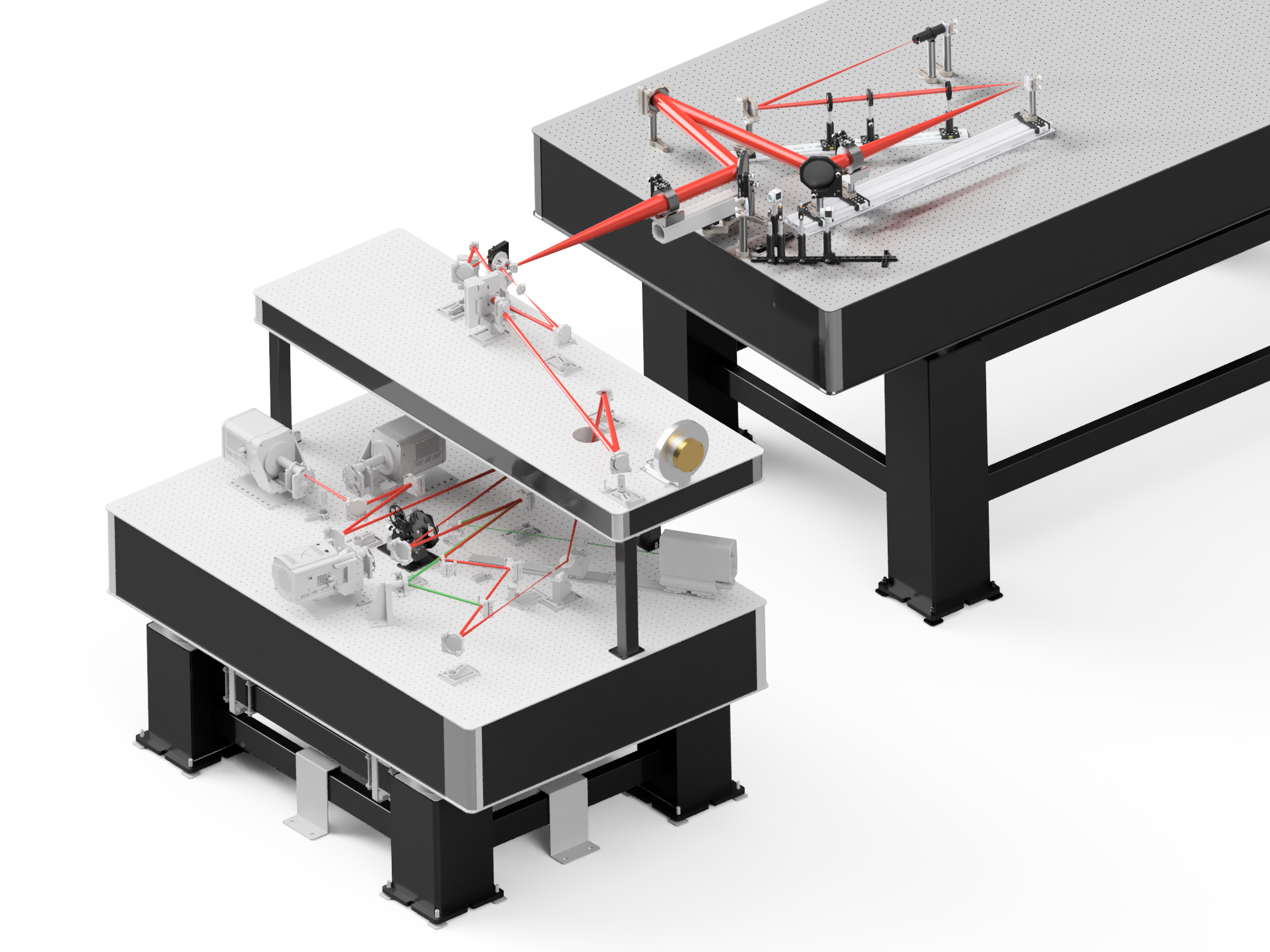}
    \\[6pt]
    \caption{A solid model rendering of p-HCAT feeding MagAO-X. Light travels through a hole in the wall (not shown here) and into MagAO-X. Both systems are built on floating optical tables but MagAO-X has a TMC\textsuperscript{TM} PEPS-II closed-loop feedback system to maintain a consistent alignment between the two tables.}
    \label{fig:feeding-magao-x}
\end{figure}

To feed light into MagAO-X, we send light through a hole in the wall into the adjacent lab, where MagAO-X is located (see Figure \ref{fig:feeding-magao-x}). Both the MagAO-X instrument and p-HCAT were built on floating optical tables, but the MagAO-X instrument has a TMC\textsuperscript{TM} PEPS-II closed-loop feedback system to lock the height of the floating table. This allows us to maintain a consistent alignment of the two tables from a day-to-day basis.

\subsection{The HDFS}
\label{sect:hdfs}

\begin{figure}[b!]
    \centering
    \includegraphics[width = 0.88\textwidth, angle = 0]{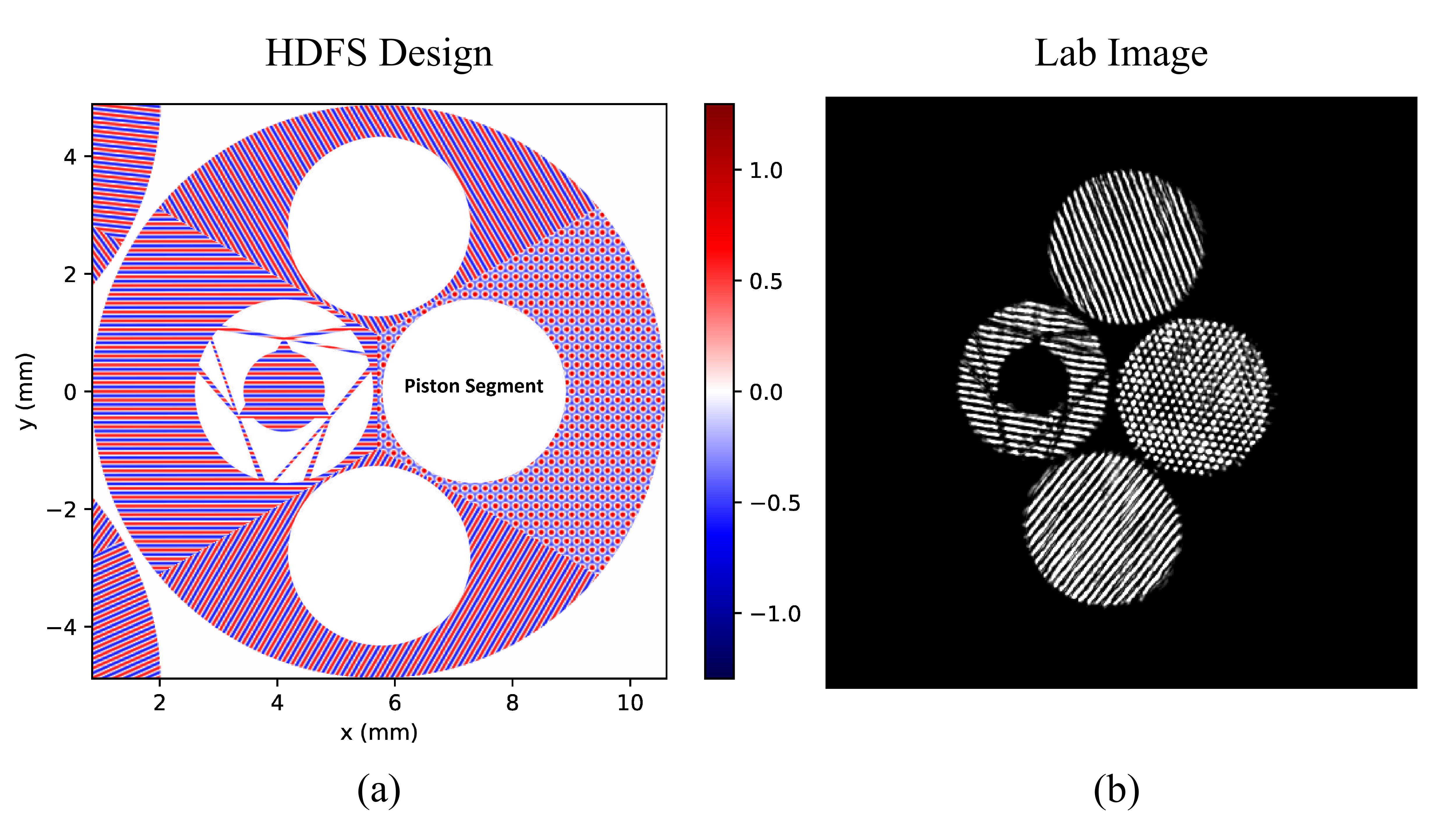}
    \\[6pt]
    \caption{(a) The HDFS design with the p-HCAT pupil. The piezo segment is on the right hand side, so the holographic phase is a multiplex of cosine patterns to interfere this segment with the other three GMT segments. In this fashion, the differential phase between the three pairs of segments can be simultaneously measured (from 6 dispersed fringes) and solved for the GMT's piston. (b) A lab image of the pupil with the HDFS placed in the pupil wheel as seen by the MagAO-X pupil viewer.}
    \label{fig:hdfs-pupil}
\end{figure}

The HDFS is a novel phase sensor that was designed as a second piston sensing channel for HCAT to sense any 2$\uppi$ phase ambiguities from the PyWFS (see Haffert et al.) \cite{haffert_2022}. This special optic is a custom continuous phase plate made of a complex liquid crystal array (frozen in resin) that has been optimized for the special geometry of the GMT pupil (see Figure \ref{fig:hdfs-pupil}). The HDFS utilizes the full GMT aperture, interfering pairs of 8.4-m segments and dispersing the light to observe dispersed fringes in the focal plane. Three different prototypes of the HDFS were fabricated by BEAM Co. for the HCAT project. One was designed for the four-segment p-HCAT pupil, while the other two were designed for the seven-segment HCAT pupil. The prototypes were designed to be placed in one of MagAO-X's pupil plane wheels (as shown in Figure \ref{fig:magao-x}) to create dispersed fringes on the science cameras. Figure \ref{fig:hdfs-pupil} shows the HDFS design and a lab image of the p-HCAT pupil with the HDFS liquid crystal optic placed in the beam as seen with MagAO-X's pupil image viewer.

\begin{figure}[b!]
    \centering
    \includegraphics[width = 0.9\textwidth, angle = 0]{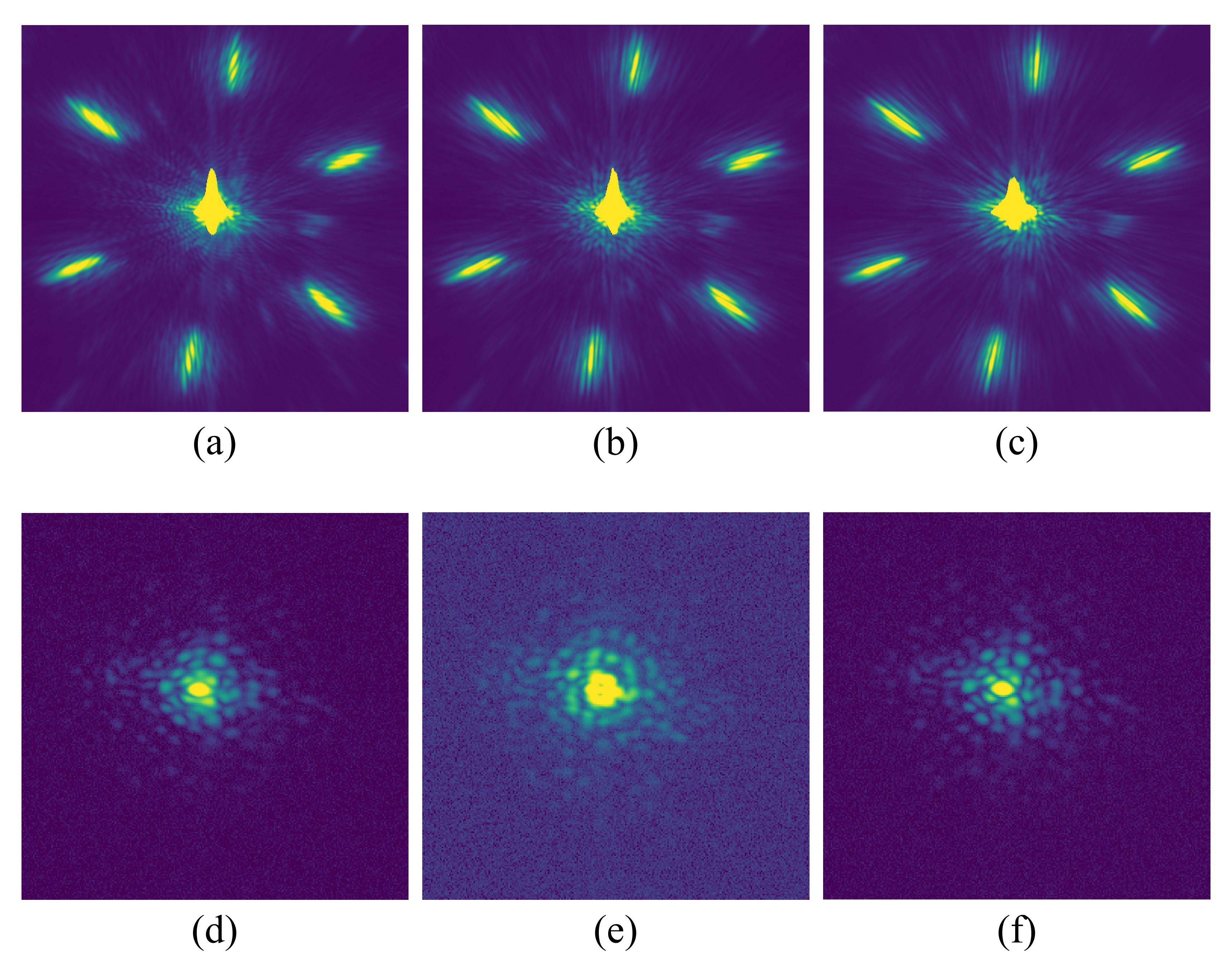}
    \\[6pt]
    \caption{Closed-loop images of the HDFS PSF and science PSF taken simultaneously with the MagAO-X science cameras. These images are part of a series of calibration images that were taken for reconstructing the piston error. (Top row): A log stretch of the HDFS PSF in broadband light (0.60\,$\upmu$m -- 0.90\,$\upmu$m) with (a) 7.0\,$\upmu$m (16$\uppi$ phase) of piston wavefront error introduced by the p-HCAT piston segment, (b) 3.0\,$\upmu$m (7$\uppi$ phase) of piston wavefront error, and (c) 0.0\,$\upmu$m (0$\uppi$\,phase) of piston wavefront error. (Bottom row): A log stretch of the science PSF viewed on a separate science camera at $\uplambda_c$ = 875\,nm, 50\,nm bandwidth with (d) 7.0\,$\upmu$m of piston wavefront error introduced by the p-HCAT piston segment, (e) 3.0\,$\upmu$m of piston wavefront error, and (f) 0.0\,$\upmu$m of piston wavefront error. The PSFs look very similar with 7.0\,$\upmu$m and 0.0\,$\upmu$m piston error because the intensity wraps by 2$\uppi$ for each wave of piston wavefront error. Therefore, it is difficult to tell that the Strehl ratio of the PSF in (d) ($\sim$\,60\% at 875\,nm) is lower than the co-phased PSF in (f) ($\sim$\,70\% at 875\,nm) because the intensity has wrapped by 16$\uppi$ (8\,$\uplambda$), so they look deceptively similar. On the other hand, the HDFS PSFs have differences in each image that are qualitatively obvious. To see how this works: the middle panel shows HDFS fringes with three maxima from 0.60\,$\upmu$m\,--\,0.90$\upmu$m. This can only occur if the piston error at 0.60\,$\upmu$m is 5\,$\uplambda$ OPD, the piston error at 0.75\,$\upmu$m is 4\,$\uplambda$ OPD, and the piston error at 0.90\,$\upmu$m is $\sim$\,3\,$\uplambda$ OPD. Each of these phase errors calculates to 3.0\,$\upmu$m of OPD, which is in perfect agreement with the piston OPD we applied. Hence, this confirms that the HDFS fringes are measuring the exact piston OPD error. This illustrates that piston sensing with only a PSF is nearly impossible while piston sensing with the HDFS is much more deterministic and linear.}
    \label{fig:hdfs-psf}
\end{figure}

Figure \ref{fig:hdfs-psf} shows closed-loop images of the HDFS PSF and the science PSF (in log stretch) taken simultaneously with the MagAO-X science cameras. These were some of the calibration images that were used as a template for reconstructing the piston error (See Section \ref{sect:hdfs-no-turbulence} and Haffert et al. \cite{haffert_2022}). The HDFS was operated on one science camera in broadband light (0.60\,$\upmu$m\,--\,0.90\,$\upmu$m) while the other science camera was used to view the PSF in one astronomical band ($\uplambda_c$\,=\,875\,nm, 50\,nm bandwidth). The left column of Figure \ref{fig:hdfs-psf} shows 7.0\,$\upmu$m (16$\uppi$ phase) of piston wavefront error introduced by the p-HCAT piezo segment, the middle column shows 3.0\,$\upmu$m (7$\uppi$ phase) of piston error, and the right column shows 0.0\,$\upmu$m (0$\uppi$ phase) of piston error (on the white light fringe of the system when co-phased). The Strehl ratio of the co-phased PSF with 0.0\,$\upmu$m piston error was measured to be $\sim$\,70\% at 875\,nm while the PSF with 7.0\,$\upmu$m wavefront error was measured to be $\sim$\,60\% at 875\,nm. These narrowband PSFs look very similar because the intensity wraps by 2$\uppi$ per 1$\uplambda$ of piston wavefront error, illustrating that piston sensing with only a PSF is nearly impossible while piston sensing with the HDFS is much more deterministic and linear.

\section{Results of Closed-Loop Piston Control with p-HCAT}
\label{sect:results}

It is well known that the sensitivity of a PyWFS can be adjusted by modulating the focal plane around the tip of the pyramid via a tip/tilt mirror in a pupil plane \cite{esposito_2000}. Zero modulation results in the highest sensitivity and smallest dynamic range, while modulation results in less sensitivity and larger dynamic range. Modulation is typically needed for an AO system to close the loop on-sky since the wavefront aberration of a seeing-limited source exceeds the dynamic range of an unmodulated PyWFS. It has also been shown that a PyWFS's response to differential piston error from the GMT segments is stronger at zero modulation and weaker as modulation increases \cite{van-dam}. For these reasons it is difficult to discern whether or not a PyWFS can truly measure and correct differential piston errors from the GMT segments in the presence of atmospheric turbulence without testing it with real optics and hardware in a lab environment. Here we report the first results of closed-loop piston control with p-HCAT and MagAO-X's PyWFS with and without turbulence for several different modulation radii. We also report the results of closed-loop piston control with the novel HDFS as a second piston sensing channel (see Haffert et al.) \cite{haffert_2022}.

\subsection{Calibrating the AO system and the PyWFS}
One of the main challenges for calibrating the AO system with p-HCAT was the low-order bench turbulence from the p-HCAT lab. Special care was taken to calibrate the PyWFS under these circumstances with the presence of a time varying disturbance. The interaction matrix was determined from a long time series of random Gaussian probes. At each step, all modes were perturbed by random Gaussian noise, and at each subsequent step the opposite random probe was applied. The difference measurement for a single step is given by,
\begin{equation}
    s_i = 2 A p_i + \Delta r_i,
\end{equation}
where $s_i$ is the measurement at step $i$ and $p_i$ is the vector containing the modal coefficients of probe $i$. The interaction matrix $A$ converts the modal coefficients into a wavefront sensor measurement. The low-order background signal changes between the positive and negative probe which introduces an offset in the measurement of $\Delta r_i$. The background signal can be averaged out over time if enough probes are used and the difference in the background signal is distributed as $\Delta r_i \sim \mathcal{N}(0, \Sigma)$, where $\Sigma$ is the covariance of the background process. The optimal least squares solution is given by,
\begin{equation}
    A =\frac{1}{2} S \Sigma^{-1}P^T(P \Sigma^{-1}P^T)^{-1}.
\end{equation}
Here, all measurements $s_i$ and probes $p_i$ have been collected and rearranged into matrices $S$ and $P$. Instead of the optimal estimator, we used the regularized normal least-squares solution, 
\begin{equation}
    A =\frac{1}{2} SP^T(PP^T + \mu I)^{-1},
\end{equation}
where $\mu$ is the regularization parameter. We assumed that the low-order modes are uncorrelated and identically distributed, so $\Sigma = \sigma I$. Under this assumption, and the assumption that $\mu$ = 0, the normal least-squares solution is the optimal least-squares solution. While this assumption may not be completely true, it does not lead to a bad interaction matrix in practice. However, it may be possible to increase the quality of the interaction matrix if knowledge of the disturbance is added. 

The modes that were calibrated for the interaction matrix were tip, tilt, the first 500 Fourier modes of the MagAO-X tweeter DM and the piston mode of the PI stage from the p-HCAT Holey Mirror. This led to a total of 503 calibrated modes. The inclusion of the piston mode as a separate mode from the PI stage during the calibration of all Fourier modes was crucial since the 2k tweeter DM is capable of reproducing piston with enough Fourier modes. Hence, if the piston mode was not explicitly added as a separate mode, it would have been corrected by the Fourier modes of the tweeter and this would not have been a valid test of piston control with a segmented telescope.

\subsection{PyWFS: No Turbulence}
\label{sect:pwfs-no-turbulence}

\begin{figure}[ht!]
    \centering
    \includegraphics[width = 0.57\textwidth, angle = 0]{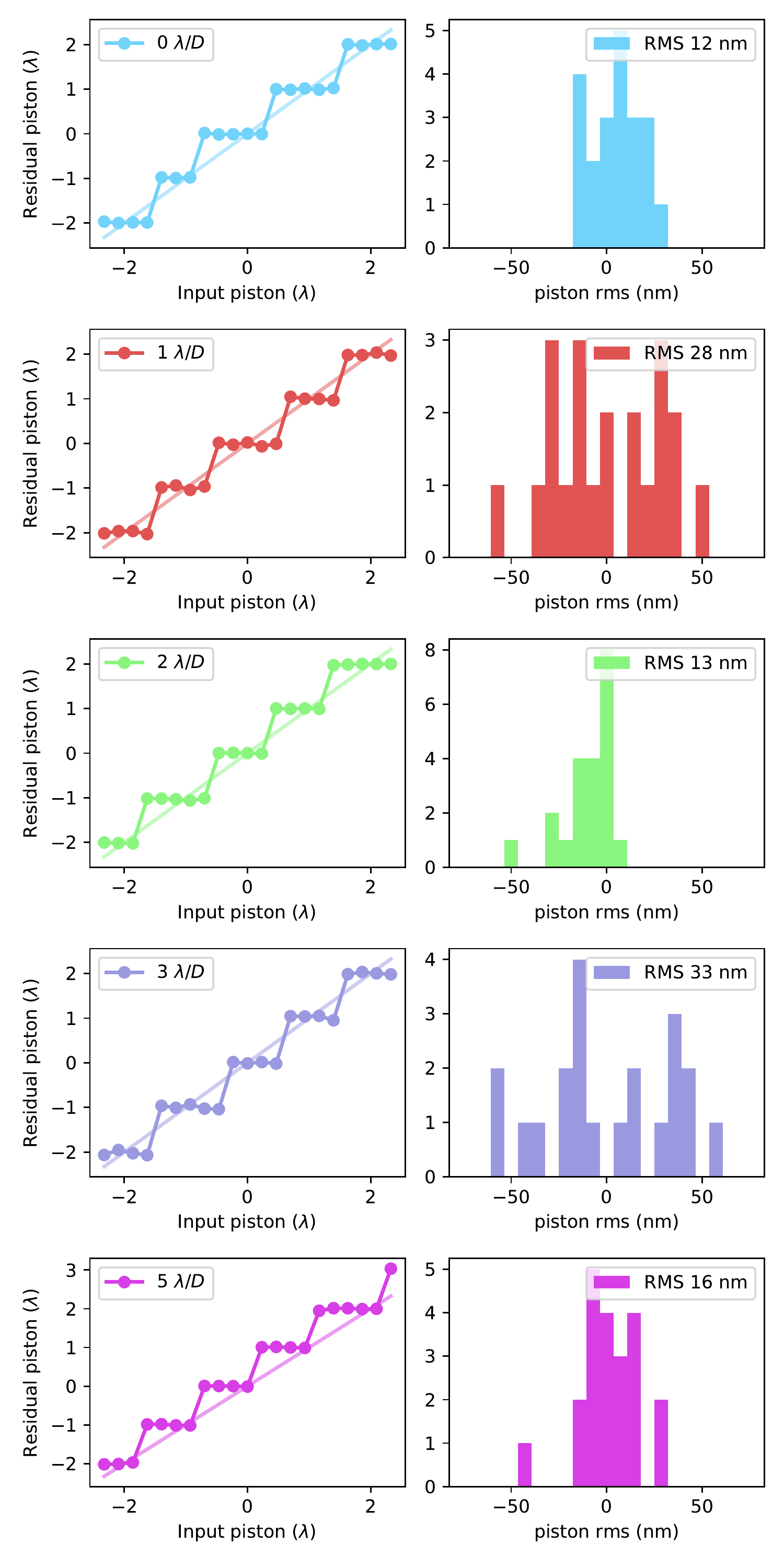}
    \\[6pt]
    \caption{The PyWFS residual piston error as a function of input piston for a variety of modulation radii. We injected consistent steps of $\uplambda$/5 OPD piston with the piezo segment from p-HCAT while closed-loop with the PyWFS and plotted the residual piston error as corrected by the PyWFS. We can see that the PyWFS wraps by 2$\uppi$ for all modulation radii, limiting the dynamic range of the PyWFS to $\pm$\,$\uplambda$/2. However, the precision of the piston corrections was quite good ($<$\,33\,nm\,RMS for all modulation radii).}
    \label{fig:pwfs-all}
\end{figure}

First we tested the PyWFS's sensitivity to piston with no turbulence at different modulation radii by inserting $\uplambda$/5 OPD steps of piston jumps with the piezo segment from p-HCAT, measuring and controlling the induced piston errors with the PyWFS in closed-loop, and plotting the residual piston error for multiple waves of injected piston. We repeated this test multiple times to calculate the precision of the PyWFS measurements. Figure \ref{fig:pwfs-all} shows the plots of the residual piston error as a function of input piston for 0\,$\uplambda$/D, 1\,$\uplambda$/D, 2\,$\uplambda$/D, 3\,$\uplambda$/D, and 5\,$\uplambda$/D modulation. 

Ideally, the residual piston error would be constant at zero for all waves of input piston error. However, since the PyWFS signal in response to wavefront phase error is sinusoidal in nature \cite{esposito_2000}, the PyWFS signal wraps by 2$\uppi$ as the injected piston approaches $\uplambda$/2 OPD. As a result, the PyWFS finds that the phase error is ``just as good" at 1\,$\uplambda$ OPD as it is at 0\,$\uplambda$ OPD, which ultimately limits the PyWFS's dynamic range to $\pm \uplambda$/2 OPD. We also found that the ``bench seeing" from the p-HCAT lab created enough residual wavefront error to make the PyWFS jump by 2$\uppi$ earlier than expected in some cases. This can be seen by the difference in the number of data points between consecutive 2$\uppi$ phase jumps in the left plots of Figure \ref{fig:pwfs-all}. The plots on the right show the histograms of the wrapped residuals, which indicate that the precision of the reconstructed piston error was within 12 -- 33\,nm\,RMS for 0$\uplambda$/D -- 5$\uplambda$/D modulation.

\begin{figure}[ht!]
    \centering
    \includegraphics[width = 1.0\textwidth, angle = 0]{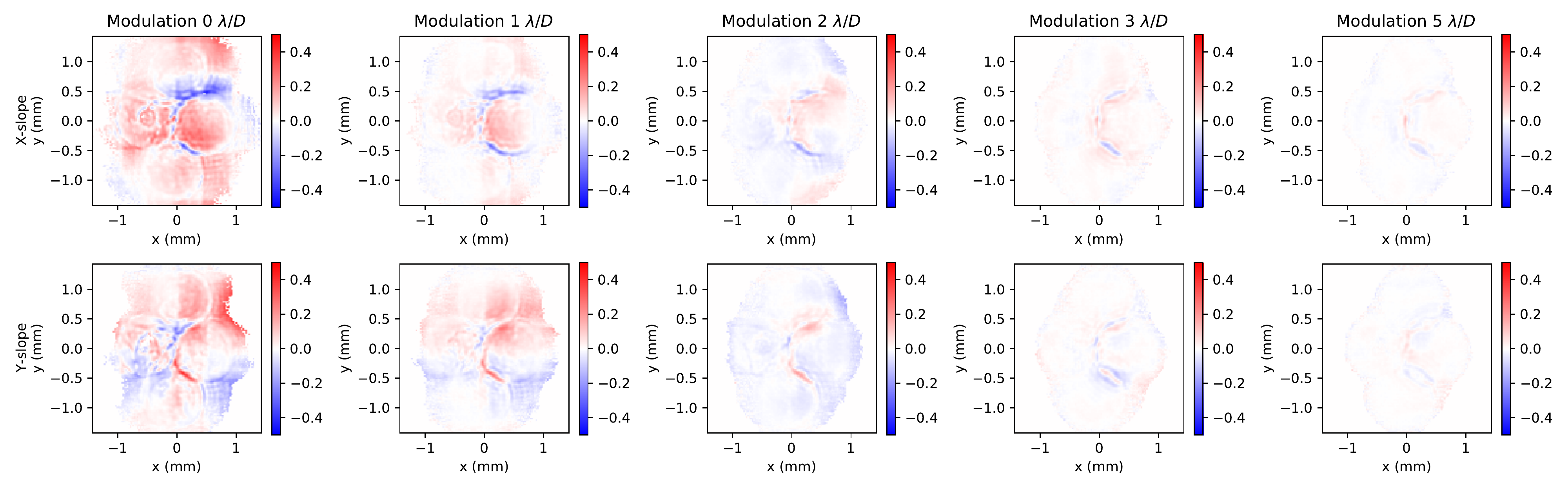}
    \\[6pt]
    \caption{The PyWFS piston slope response as a function of modulation radius with $\pm$\,100\,nm piston (wavefront). The top row shows the $x$ slope and the bottom row shows the $y$ slope. As the modulation radius approaches 5\,$\uplambda$/D, the PyWFS's sensitivity to piston decreases.}
    \label{fig:pwfs-slopes}
\end{figure}

Figure \ref{fig:pwfs-slopes} shows the measured PyWFS slope response for each modulation radius with $\pm$\,100\,nm of piston error (wavefront). As the modulation increases, the PyWFS becomes less sensitive to piston, however we were still able to control piston up to 5\,$\uplambda$/D modulation. These results were better than predicted, but at 5\,$\uplambda$/D we were more sensitive to low flux conditions. From these tests we can conclude that the PyWFS has the ability to measure and correct piston errors from 0\,$\uplambda$/D -- 5\,$\uplambda$/D modulation to within 12 -- 33\,nm\,RMS with no turbulence, but has a limited dynamic range ($\pm \uplambda$/2 piston wavefront error).

\subsection{PyWFS: With Turbulence}
\label{sect:pwfs-turbulence}

The next stage of our tests was to inject a Kolmogorov turbulence phase screen with MagAO-X's 2,040 actuator tweeter DM and close the loop with the PyWFS and the piezo segment. To make sure these tests were accurate, the response matrix was updated and tested successfully without turbulence first. We applied two different turbulence phase screens which correspond to 0.6 arcsec (median seeing conditions at the GMT site) \cite{thomas-osip-2008} and 1.2 arcsec seeing at 500\,nm, then we closed the loop starting with zero piston error introduced by the piezo segment. 
\begin{figure}[ht!]
    \centering
    \includegraphics[width = 1.0\textwidth, angle = 0]{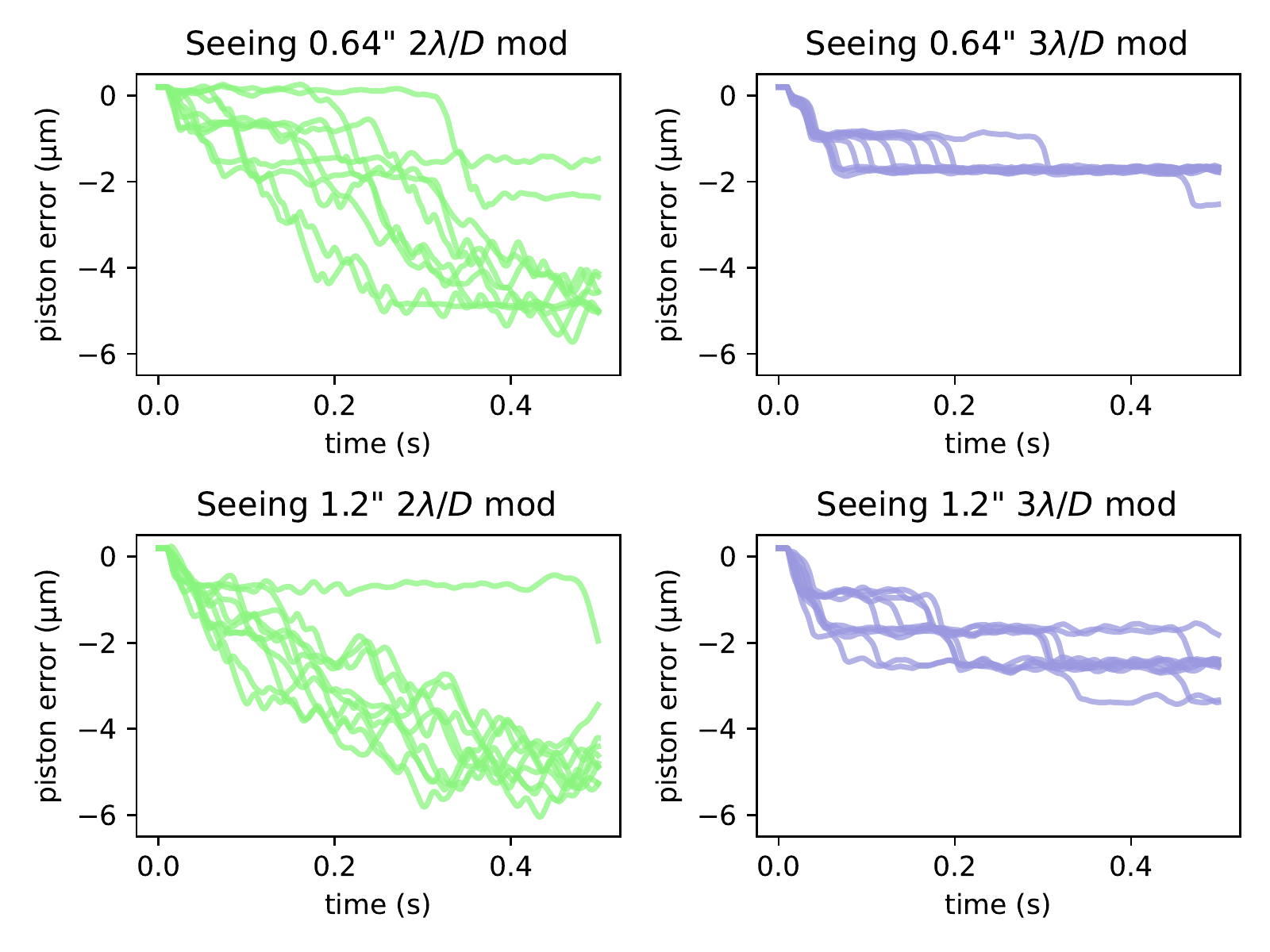}
    \\[6pt]
    \caption{The PyWFS residual piston error as a function of time for 2\,$\uplambda$/D and 3\,$\uplambda$/D modulation with 0.6 arcsec seeing and 1.2 arcsec seeing. The piston error introduced with the p-HCAT piezo segment started at zero piston error for each iteration and the PyWFS always failed to keep the segment phased. For each case, the PyWFS continuously ``ran away" in piston, likely due to non-linear cross talk between modes and poor pixel sampling of the gaps between segments on the PyWFS detector. The loop was more stable with 3$\uplambda$/D modulation than 2$\uplambda$/D modulation because the PyWFS's sensitivity decreases and the linearity increases with modulation, reducing the non-linear cross-talk effect.}
    \label{fig:pwfs-atmosphere}
\end{figure}
Figure \ref{fig:pwfs-atmosphere} shows the reconstructed piston error as a function of time for 2\,$\uplambda$/D and 3\,$\uplambda$/D modulation with 0.6 arcsec seeing and 1.2 arcsec seeing. For each case, the PyWFS quickly detected piston phase errors outside of its dynamic range and continuously ``jumped" by 2$\uppi$. We performed this test for multiple iterations and different modulation radii, but we could not find any value of modulation (0\,$\uplambda$/D, 1\,$\uplambda$/D, 2\,$\uplambda$/D, 3\,$\uplambda$/D, or 5\,$\uplambda$/D) or gain value of the PyWFS that allowed a stable piston control for more than one minute. The results were the same: the PyWFS continuously ``ran away" in piston and failed to close the loop with the piezo segment. 

One reason why we may have experienced this phenomenon of the PyWFS ``running away" in piston due to atmospheric turbulence is explained in detail by Bertrou-Cantou et al. (2022) \cite{bertrou_2021}. For a segmented aperture like the GMT, TMT or the ELT, which has $\sim$\,50\,cm gaps that span multiple atmospheric coherence lengths in the visible, there will be a difference in the average phase value of the turbulence across each segment, referred to as a ``petal." These large gaps between segments will cause a discontinuity in the fitted wavefront data that a PyWFS will see as a differential piston error that needs to be corrected. Bertrou-Cantou et al. show that when these petal modes are incorporated into the interaction matrix of an AO system, the PyWFS will experience a heavy non-linear cross-talk between the petal modes and high-order modes that is dependent on turbulence conditions, making it very hard to calibrate with optical gain (OG) compensation. In theory, a high fidelity OG compensation method could be used to resolve this issue, but it is very likely that there will be some error in OG measurements, especially in harsh seeing conditions ($\sim$\,1\,arcsec) that will push the PyWFS out of its minimal dynamic range and cause the PyWFS to continuously phase wrap by 2$\uppi$ and ``run away" in piston. 

\begin{figure}[ht!]
    \includegraphics[width = 0.9\textwidth, angle = 0]{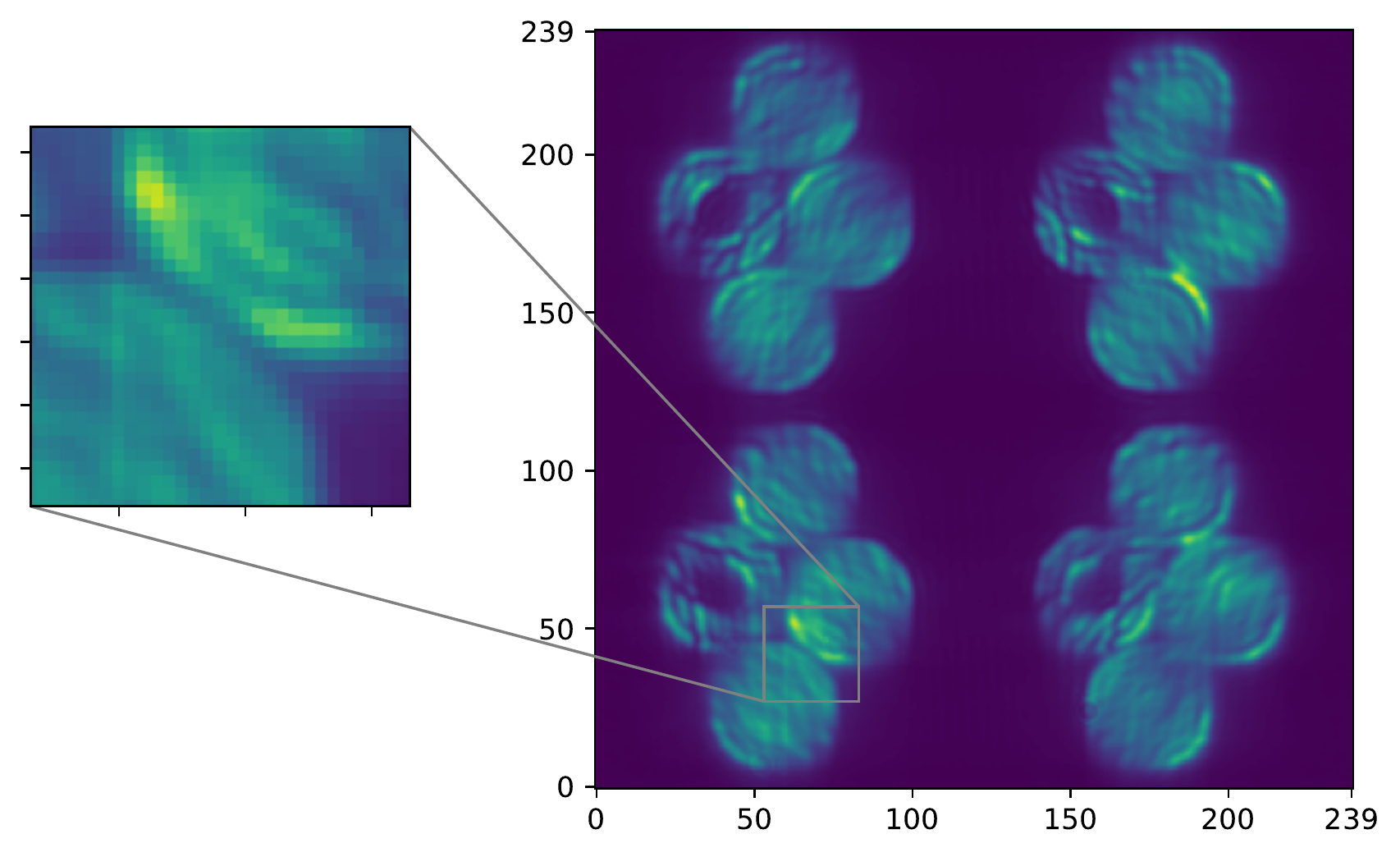} 
    \caption{The MagAO-X PyWFS signal with 3\,$\uplambda$/D modulation and full OCAM2k sampling (bin\,1 240\,x\,240 pixels). There is zero piston error introduced by the p-HCAT piezo segment and the wavefront was flattened with the MagAO-X 2,040 actuator tweeter DM. We can see that even with no piston error and a flat wavefront, there is still some signal in the gaps. There is only 1 pixel in between each segment, so it is very easy to scatter or diffuse light into the gaps. At lower sampling, this effect will become even stronger. Unfortunately, it is impossible to achieve better GMT pupil sampling with today's highest speed (2\,kHz frame rate) EMCCD technology.}
    \label{fig:pupil-sampling}
\end{figure}

There are also practical reasons why a PyWFS is difficult to use in turbulence. Any light that leaks (or scatters) in the gaps between the segments on the PyWFS detector will cause the PyWFS to sense a piston error, since this is the region where the strongest piston signal appears (as shown in Figure\,\ref{fig:pwfs-slopes}) \cite{van-dam}. For this reason it is important to have sufficient pixel sampling in the segment gaps.  Figure \ref{fig:pupil-sampling} shows an image of the MagAO-X PyWFS signal with the p-HCAT pupil. We have no more than $\sim$\,1 totally isolated EMCCD pixels between the segment gaps, so any light that leaks into this region due to CCD charge diffusion (from bright high-order wavefront slopes on either side of the gap) or split frame transfer bleed, etc. may convince the PyWFS that there is a piston signal and cause the PyWFS to ``run away" in piston.

\subsection{HDFS: No Turbulence}
\label{sect:hdfs-no-turbulence}

There is a clear need for a second piston sensing channel that can either catch 2$\uppi$ phase jumps from the PyWFS or take complete control over piston sensing. We have explored the use of the novel HDFS as a second piston channel for this purpose. First we tested the HDFS with no turbulence and successfully demonstrated closed-loop piston control with the piezo segment from p-HCAT down to 20 nm RMS (see Haffert et al. for more details on these results) \cite{haffert_2022}. The PyWFS was used purely as a slope sensor here (piston mode gain set to zero) to correct for the low-order bench seeing and stabilize the HDFS image in the focal plane. The stabilized HDFS allowed us to take a series of calibration images by driving the piezo actuator in consecutive steps and recording the images in the focal plane. The feedback gain of the piston mode needed to be set to zero for the PyWFS since the bench seeing from the p-HCAT lab created enough differential piston to make the PyWFS jump by 2$\uppi$ earlier than expected in closed-loop (as seen in Section \ref{sect:pwfs-no-turbulence}), which interfered with the HDFS calibrations. By nullifying the piston mode from the PyWFS, we were able to close the loop without any 2$\uppi$ phase jumps and take a stable series of calibration images with the HDFS. These calibration images were used as a template library. We then applied a random amount of piston to offset the system and used the template library to measure and remove the applied piston in closed-loop, while the PyWFS was also operating in closed-loop. In this manner, the PyWFS and the HDFS worked well together to control any amount of injected piston from the piezo segment ($\pm$\,10\,$\upmu$m total range) to within 20\,nm\,RMS \cite{haffert_2022}. 

\subsection{HDFS: With Turbulence}
\label{sect:hdfs-turbulence}

We performed the same test while generating 0.6 arcsec seeing turbulence (median seeing conditions at the GMT site) \cite{thomas-osip-2008} with the MagAO-X tweeter DM. The PyWFS was used purely as a slope sensor again with the piston mode feedback gain set to zero, while the HDFS used its calibration images to close the loop with the piezo segment. Removing the piston mode from the PyWFS's interaction matrix allowed us to completely bypass any unwanted 2$\pi$ phase ambiguities experienced in Section \ref{sect:pwfs-turbulence} so that the HDFS could take complete control of piston. After generating turbulence with the tweeter DM and closing the loop with the PyWFS, we injected random amounts of piston for multiple iterations and closed the loop with the HDFS. In this way the HDFS was able to successfully close the loop with the piezo segment from p-HCAT for both seeing conditions to within 50\,nm\,RMS (see Haffert et al.) \cite{haffert_2022}.

There is one problem we faced during these tests that limited our results. Since the MagAO-X science cameras could only run as fast as 20\,Hz, we could only run the HDFS at 20\,Hz, which was too slow to keep up with the simulated atmosphere that was injected at 300 Hz. To resolve this issue, the atmosphere was ``slowed down," meaning the loop was updated at a slower rate than the simulated atmosphere. The slower the relative loop speed, the faster the effective speed of the HDFS. However, since the HDFS was calibrated at 20\,Hz while the PyWFS was controlling tip/tilt at the full frame rate of the WFS detector (300\,Hz) to control the bench seeing from the p-HCAT lab, residual bench seeing dominated the reconstruction errors when slowing down the atmosphere too much. Hence, a balance was found at a relative loop speed of 60\,Hz, where the HDFS reached the best correction error of 50\,nm\,RMS. 

This problem could be eliminated by simply using a faster camera to allow the HDFS to keep up with the atmosphere and fix the bench seeing errors. In Haffert et al.\cite{haffert_2022} we demonstrate the HDFS working in closed-loop without p-HCAT (internally inside MagAO-X without any bench seeing) to reach a correction error of $\sim$\,5\,nm\,RMS. Therefore, we expect to reach this same level of precision when we incorporate a faster camera for the next phase of the HCAT project.

\subsection{Discussion}

It is clear from Section \ref{sect:pwfs-no-turbulence} and \ref{sect:pwfs-turbulence} that the PyWFS, although an ideal wavefront sensor for traditional AO, is not a good piston sensor for GMT, ELT, and TMT ExAO, since the PyWFS will run away in piston due to the sinusoidal nature of the PyWFS signal in response to piston and non-linear modal cross-talk between segment/petal modes and high-order modes that depends on atmospheric turbulence conditions. Also, any light that leaks between the gaps of M1 segments on the PyWFS detector due to anything other than petal modes (CCD charge diffusion from high-order wavefront slopes, split frame transfer bleed, etc.) will be enough to push the PyWFS outside of its $\pm$\,$\uplambda$/2 dynamic range, causing the PyWFS to converge to multiples of 2$\uppi$ and continuously run away in piston. For this reason, GMT, ELT, and TMT ExAO instruments will need a second piston sensing channel (such as the HDFS) to control segment/petal modes alone while the PyWFS works purely as a slope sensor with segment/petal modes nullified from its interaction matrix.

The HDFS does not have the issue of running away in piston since it sees multiple wavelengths all at once by creating dispersed fringes in the focal plane. In addition, the HDFS utilizes each 8.4-m segment of the GMT, creating a high signal-to-noise ratio for piston sensing with faint stars. Numerical simulations show that the HDFS running at 1\,kHz only needs 10\% of the light from a 12\textsuperscript{th} magnitude star in J and H band (1.1\,$\upmu$m\,--\,1.8\,$\upmu$m) to control piston to within 50\,nm\,RMS (see Haffert et al.) \cite{haffert_2022}. The only caveat is that the HDFS needs a stable PSF in the focal plane created with an AO system. Hence, a PyWFS used purely as a slope sensor (with segment/petal modes nullified) is the perfect partner for the HDFS. 


\section{HCAT}
\label{sect:HCAT}

The second stage of the HCAT project is the full seven segment GMT phasing testbed. HCAT is an upgraded version of p-HCAT that simulates all seven GMT segments with six segments that can piston, tip, and tilt using PI S-325 actuators. One of the main goals of HCAT is to test a concept for the GMagAO-X ``parallel DM" that splits the pupil onto seven commercially available DMs using a reflective hexagonal pyramid \cite{close_gmagaox}. Here we discuss the parallel DM concept and the current design of HCAT.

\subsection{GMagAO-X and the Parallel DM}
\label{sect:parallel-dm}
GMagAO-X is a potential ExAO system and exoplanet imager for GMT high-contrast NGSAO science \cite{males_gmagaox, close_gmagaox}. It has just passed its GMT Conceptual Design Review (CoDR) in September 2021. As motivated in Males et al. (2019) \cite{males_gmagaox} and Close et al. (2019) \cite{close_gmagaox}, using wavelengths as blue as 0.70\,$\upmu$m is optimal to characterize low-mass, temperate exoplanets in reflected light. For an ExAO coronagraph to function at these short wavelengths, the fitting error of the wavefront correction must be minimized to an acceptable level ($<$\,50\,nm RMS). Therefore, it is necessary to have a massive ($>$\,20,000 actuator) DM for any GMT/ELT/TMT ExAO system for exoplanet science. The issue is that no commercially available DM anywhere near this size exists today. Hence, we need to ``parallelize" the problem into smaller parts that can be handled by commercial DMs. This is the GMagAO-X ``parallel DM" concept \cite{close_gmagaox}. 

\begin{figure}[ht!]
    \centering
    \includegraphics[width = 0.95\textwidth, angle = 0]{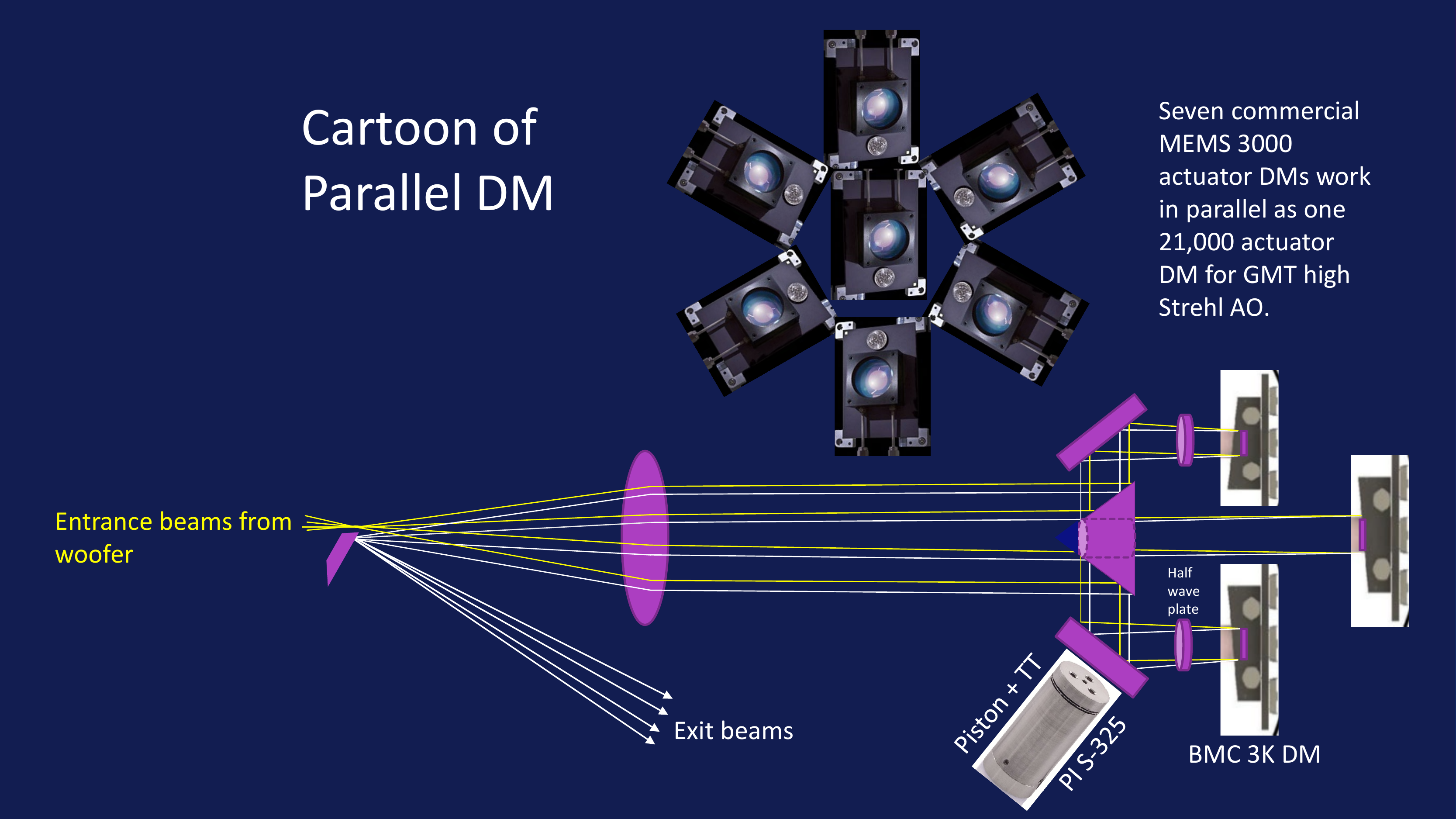}
    \\[6pt]
    \caption{The GMagAO-X parallel DM concept. A reflective hexagonal pyramid splits the GMT pupil onto seven commercial BMC 3k DMs. PI S-325 actuators act as an interferometric beam combiner to control piston ($\pm$\,60\,$\upmu$m OPD) and segment tilt ($\pm$\,8\,mrad) at speeds up to 100\,Hz. After wavefront and co-phasing correction, the light is reflected back with a small global 0.14$^\circ$ angle and picked off by a knife-edge mirror to send light through the rest of the system.}
    \label{fig:parallel-dm-cartoon}
\end{figure}

In order to create a 21,000 actuator ELT-scale ExAO tweeter DM, GMagAO-X uses a reflective hexagonal pyramid to ``slice" up the GMT pupil and optically distribute each segment onto seven commercially available Boston Micromachines (BMC) 3,000 actuator DMs (see Figure \ref{fig:parallel-dm-cartoon}) \cite{close_gmagaox}. The parallel DM incorporates six PI S-325 actuators which act as an interferometric beam combiner to control piston ($\pm$\,60\,$\upmu$m OPD) and tip/tilt ($\pm$\,8\,mrad) for each off-axis segment. After wavefront and co-phasing correction, a small global 0.14$^\circ$ tilt will be applied to each DM to reflect light back towards a knife-edge mirror that will pick off the beam and send light through the rest of the system. This optical design yields a 100\% unvignetted field of view (FOV) of 3\,arcsec\,x\,6\,arcsec on-sky with the GMT and GMagAO-X.

HCAT is designed to test this concept using a prototype reflective hexagonal pyramid (see Section \ref{sect:hexpyramid}), six PI S-325 actuators, and flat mirrors as placeholders for the seven 3k DMs. A complete mounting structure made of Invar will be incorporated in the HCAT design to mount the parallel\,DM.

\subsubsection{The Hexpyramid}
\label{sect:hexpyramid}
\begin{figure}[b!]
    \centering
    \includegraphics[width = 1.0\textwidth, angle = 0]{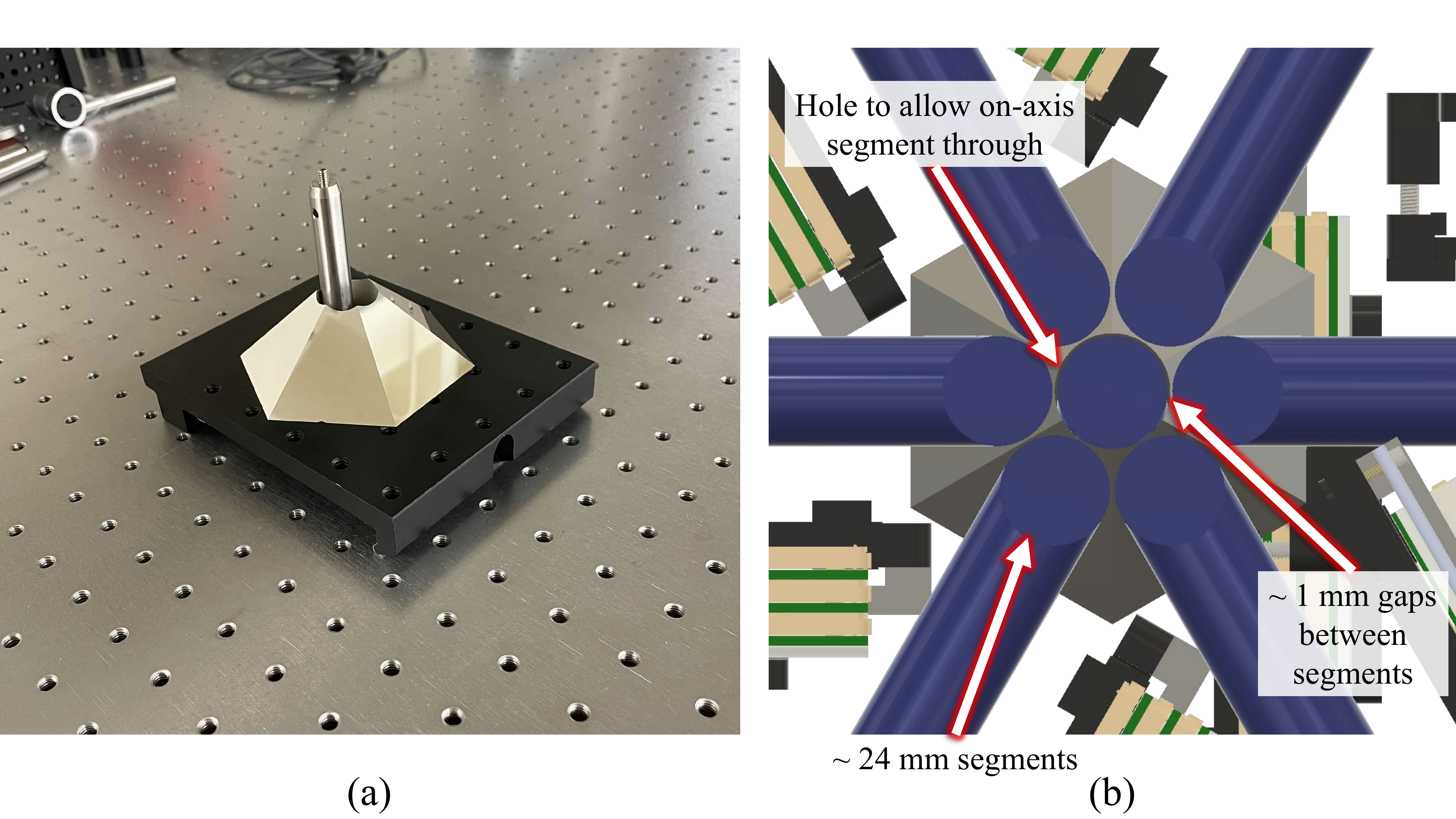}
    \\[6pt]
    \caption{(a) The hexpyramid prototype in the lab. (b) A screenshot of the solid model design. A hole in the center of the pyramid allows the on-axis GMT segment through (see Figure \ref{fig:parallel-dm}).}
    \label{fig:hexpyramid}
\end{figure}

\begin{figure}[ht!]
    \centering
    \includegraphics[width = 1.0\textwidth, angle = 0]{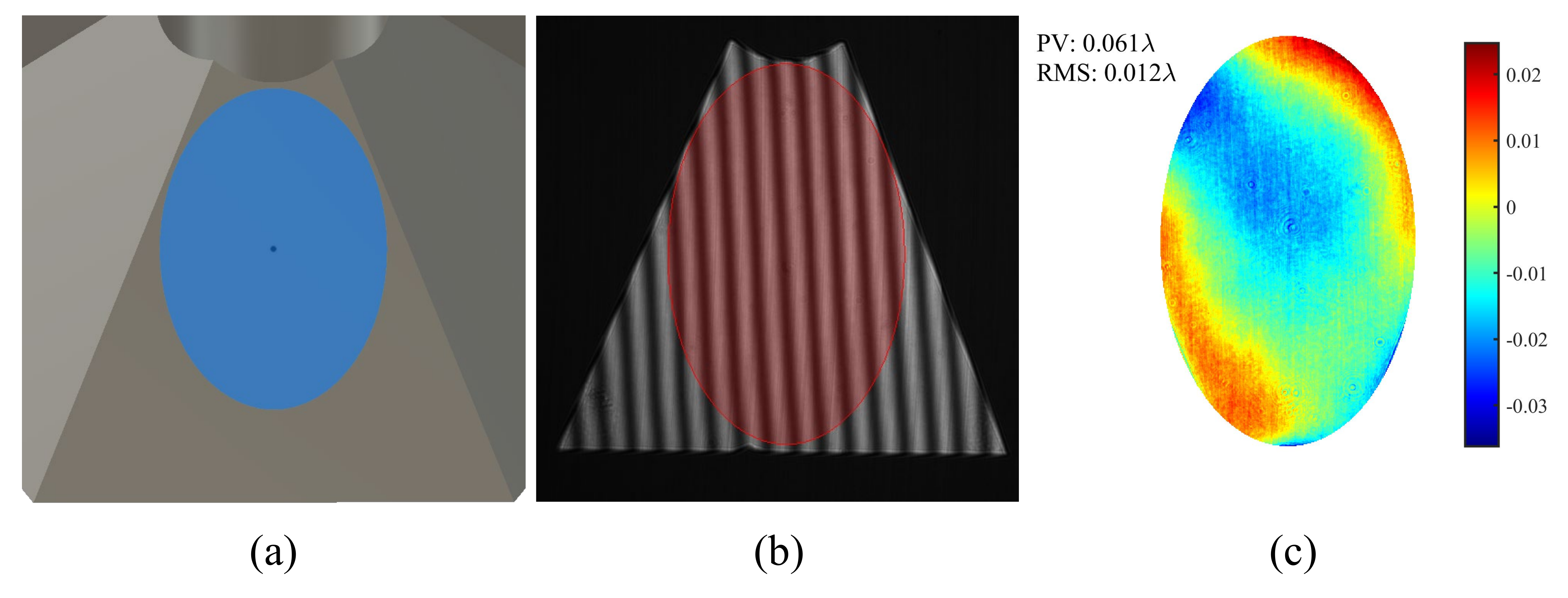}
    \\[6pt]
    \caption{(a) The beam footprint of one of the GMT segments on the hexpyramid. (b) The mask applied to the hexpyramid for the Zygo\textregistered{} measurements. (c) The Zygo\textregistered{} measurements of one of the hexpyramid faces with piston, tip, and tilt removed. All faces of the hexpyramid were measured to be $<$\,$\uplambda$/10 PV surface irregularity within the clear aperture, which is in line with our specifications. We also note that the mask applied for these measurements is larger than the actual beam footprint, so the wavefront error for each segment is expected to be even better than the Zygo\textregistered{} measurements.}
    \label{fig:hexpyramid-footprint}
\end{figure}

A prototype of the reflective hexagonal pyramid (i.e., the ``hexpyramid") was designed by Maggie Kautz and fabricated by Rocky Mountain Instrument Co. We have received the prototype in the lab and measured its surface quality with a Zygo\textregistered{} interferometer. Each surface was specified to have a protected silver coating with $< \uplambda$/10 PV surface irregularity. Figure \ref{fig:hexpyramid} shows the hexpyramid prototype in the lab and a screenshot of the solid model design. Figure \ref{fig:hexpyramid-footprint} shows the beam footprint of the off-axis segments on the hexpyramid,
the mask applied for each Zygo\textregistered{} surface measurement, and one of the Zygo\textregistered{} surface measurements. All surfaces were measured to be $< \uplambda$/10 PV over the clear aperture which is in line with our requirements. We can also see that the mask applied is larger than the beam footprint, so we expect the actual wavefront error that each segment will pick up will be even less than what we measured.

\subsubsection{Polarization Aberrations}
\label{sect:polarization-aberrations}

One of the main concerns with the parallel DM concept is the possibility of introducing polarization aberrations \cite{breckinridge_2015}. Each segment reflects at a 45$^\circ$ angle of incidence in different directions, so if the hexagonal pyramid is coated with a metallic coating like protected silver, its complex refractive index will create a phase shift between s- and p-polarization components that is incoherent between each segment, resulting in an incoherent sum of PSFs that drastically reduces the Strehl ratio. 

\begin{figure}[b!]
    \centering
    \includegraphics[width = 1.0\textwidth, angle = 0]{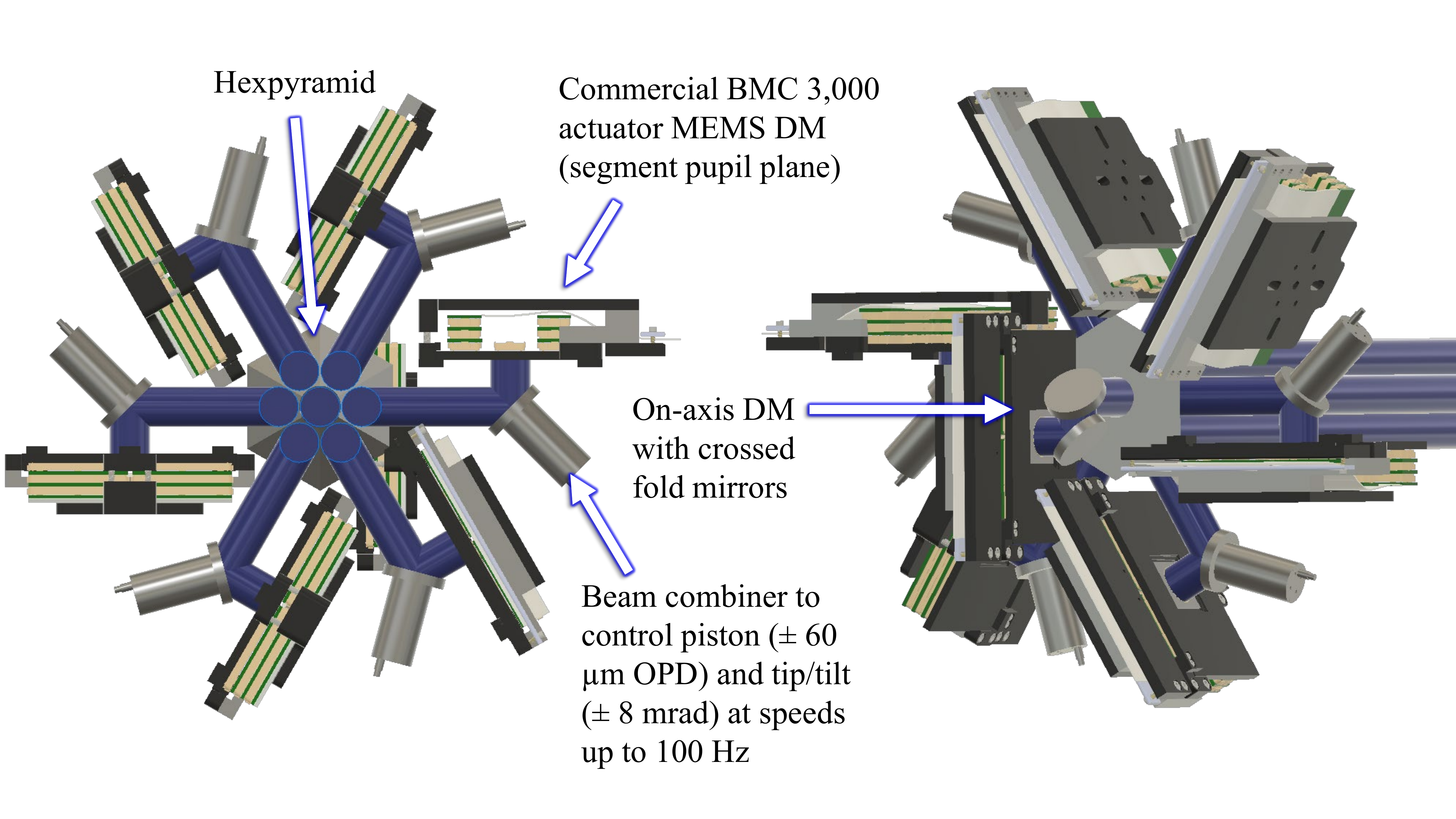}
    \\[6pt]
    \caption{The 21,000 actuator GMagAO-X parallel DM design. A reflective hexagonal pyramid splits the GMT pupil onto seven commercial BMC 3k DMs with six PI S-325 actuators to control piston, tip, and tilt for the off-axis segments. The actuated mirrors are oriented in a crossed-fold mirror configuration to minimize polarization aberrations.}
    \label{fig:parallel-dm}
\end{figure}

\begin{figure}[h!]
    \centering
    \begin{tabular}{cc}
        \includegraphics[page = 1, width = 0.45\textwidth, angle = 0]{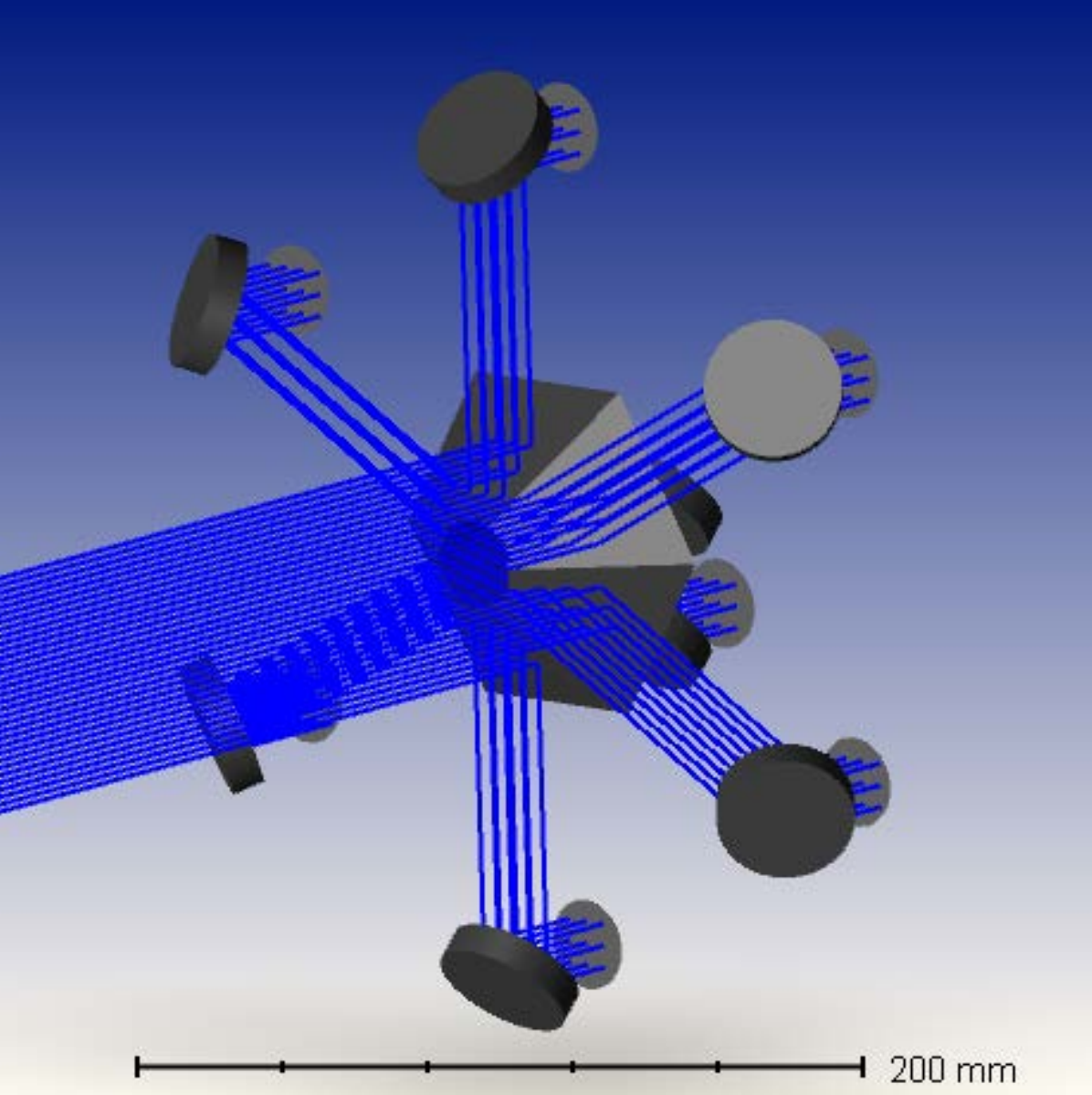} &
        \includegraphics[page = 2, width = 0.45\textwidth, angle = 0]{Figure23.pdf} \\
        (a) & (b) \\
        \includegraphics[page = 3, width = 0.47\textwidth, angle = 0]{Figure23.pdf} &
        \includegraphics[page = 4, width = 0.47\textwidth, angle = 0]{Figure23.pdf} \\
        (c) & (d) \\
    \end{tabular}
    \\[6pt]
    \caption{Zemax polarization simulation of the parallel DM (a) without crossed fold mirrors and (b) with crossed fold mirrors. (c) Without the crossed fold mirrors, the Strehl ratio varies drastically with wavelength. (d) With crossed fold mirrors, the Strehl ratio is consistent across all wavelengths.}
    \label{fig:polarization-aberrations}
\end{figure}

We have mitigated this issue by incorporating crossed fold mirrors in the parallel DM design (see Figure \ref{fig:parallel-dm}) \cite{lam_2015}. By introducing a fold mirror that is perpendicular to the plane of incidence of the hexagonal pyramid, s-polarization on the first surface becomes p-polarized on the second surface, and vice versa, so the net phase of s- and p-polarization becomes equal and the polarization aberrations perfectly cancel for the chief ray and on-axis sources, while almost cancelling perfectly for off-axis rays. For coronagraphic systems like GMagAO-X that have a small field of view, this technique is sufficient. A polarization raytrace was performed in Zemax to confirm this theory. Figure \ref{fig:polarization-aberrations} shows the Zemax results of comparing the parallel DM with and without crossed fold mirrors. A protected silver coating was applied to each mirror surface and an unpolarized light source was simulated using two orthogonal linearly polarized states. The resulting Strehl ratio as a function of wavelength is shown for each case.

We have found that the crossed fold mirror technique is sufficient for mitigating the polarization aberrations of the parallel DM. However, each mirror surface must have the same exact coating in order for the polarization aberrations to cancel. For this reason, all the fold mirrors for the HCAT parallel DM concept have been coated by the same vendor as the hexpyramid to ensure that they have the same coating.


\subsection{HCAT Optical Design}
\label{sect:optical-design-hcat}

\begin{figure}[t!]
    \centering
    \includegraphics[width = 1.0\textwidth, angle = 0]{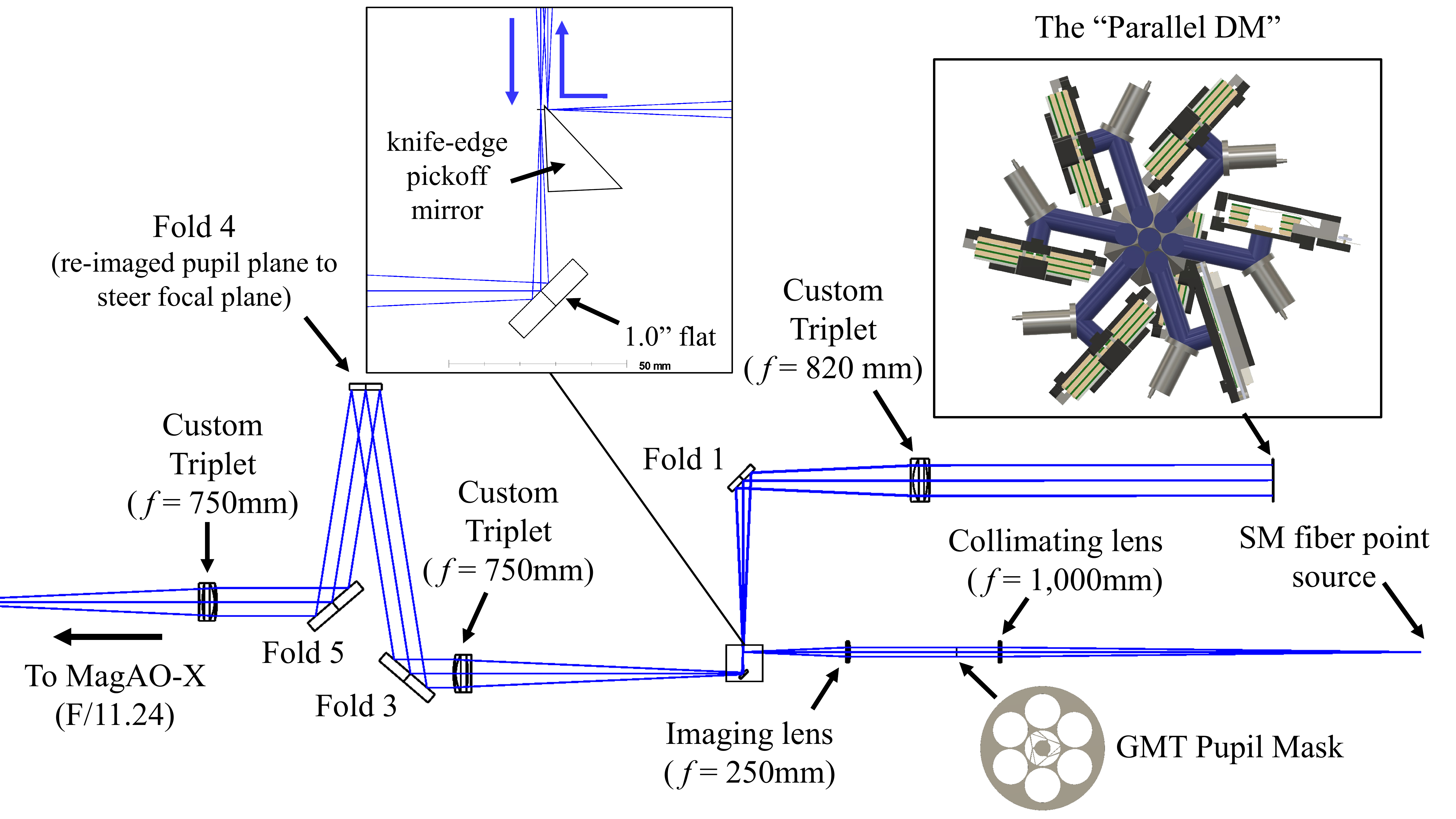}
    \\[6pt]
    \caption{The optical design of HCAT. The design consists of the same optics from p-HCAT with a third pupil relay added in the middle for the parallel DM. An Edmund Optics knife-edge mirror is placed in the first focal plane to fold the light towards the parallel DM relay. Each DM (represented by fold flats) will be tilted slightly to allow the beam to travel back to the other side of the knife-edge mirror, which folds the light towards the rest of the optics and into MagAO-X.}
    \label{fig:hcat-optical-design}
\end{figure}

The current optical design of HCAT is shown in Figure \ref{fig:hcat-optical-design}. The design uses the same optics from p-HCAT, but with a third pupil relay added in the middle for the parallel DM. A knife-edge mirror placed near the first focal plane folds the light towards a new custom triplet lens that is designed to create an image of the pupil mask with 24\,mm pupil sizes (the same size of the BMC 3k DMs). Light will reflect through the parallel DM and travel backwards in a double-pass configuration, where each ``DM" (represented by fold flats) is tilted slightly to offset the beam on the other side of the knife-edge mirror and allow the beam to travel through the rest of the system and into MagAO-X. 

\subsection{Feeding the NGWS-P}
\label{sect:feeding-ngws}

\begin{figure}[t!]
    \centering
    \includegraphics[width = 1.0\textwidth, angle = 0]{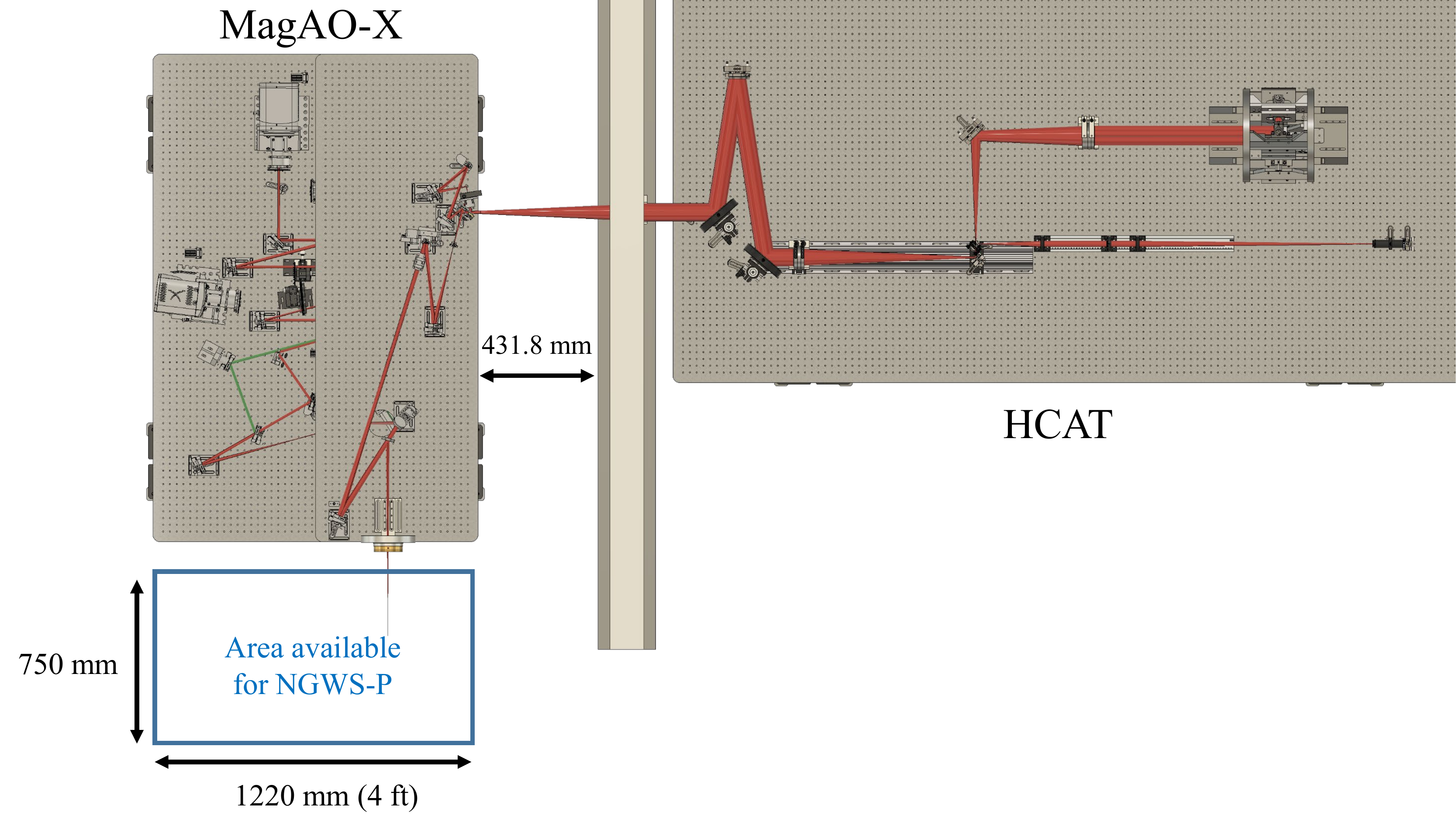}
    \\[6pt]
    \caption{The current plans to feed light into NGWS-P in 2023. The beam will be intercepted by a beamsplitter inside the MagAO-X system to send light to NGWS-P.}
    \label{fig:hcat-ngws-feed}
\end{figure}

The final stage of the HCAT project (stage 3) will be to feed light from HCAT into the prototype Natural Guide Star Wavefront Sensor (NGWS-P), which will be the official phasing testbed for the NGWS system \cite{pinna_2014}. The current plans for feeding light from HCAT into NGWS-P are shown in Figure \ref{fig:hcat-ngws-feed}. We will introduce a beamsplitter in the MagAO-X converging F/57 beam that will send light through the eyepiece port and into NGWS-P.

This is an important part of the HCAT project as it will allow us to test the real GMT NGWS internal HDFS and PyWFS hardware and control algorithms for NGSAO wavefront sensing with the GMT. The stage 1 p-HCAT tests (completed and reported in this paper) and stage 2 HCAT tests in Q3 2022 will help pave the way for stage 3 in Q2 2023, where a final robust NGWS piston sensor design will be well-tested and established with the goal of officially retiring the GMT high-risk item of phasing performance.

\section{Conclusions}
\label{sect:conclusions}

In this paper, we presented the first results of closed-loop piston control with one GMT segment from the p-HCAT testbed using MagAO-X's PyWFS and a novel Holographic Dispersed Fringe Sensor (HDFS). We showed that the PyWFS was able to control piston without turbulence for 0\,$\uplambda$/D\,--\,5\,$\uplambda$/D modulation with $\pm$\,$\uplambda$/2 dynamic range and 12-33\,nm\,RMS precision. However, with 0.6 arcsec and 1.2 arcsec seeing generated turbulence, the PyWFS failed to control piston for all modulation radii due to non-linear cross-coupling between piston and other modes as well as poor pixel sampling of the segment gaps on the PyWFS detector. We introduced a new method of sensing piston with the novel HDFS working as a ``second channel" piston sensor while the PyWFS works purely as a slope sensor and demonstrated successful closed-loop piston control to within 50\,nm\,RMS in 0.6\,arcsec atmospheric seeing. From these tests we conclude that the combination of a PyWFS as a slope sensor with a HDFS as a second channel piston sensor is a powerful architecture for AO on the GMT (and likely the TMT and ELT as well). Hence, this new HDFS + PyWFS piston sensing architecture is now the offical phasing strategy for the GMT NGWS WFS. We discussed the current design plans for the next stage of the HCAT project, which will simulate all seven GMT segments with six segments that can piston, tip, and tilt while also demonstrating a working concept for the GMagAO-X ``parallel DM" using a reflective hexagonal pyramid to slice the GMT pupil onto seven commercially-available 3k DMs. In the final section, we briefly discussed the current design plans for stage 3 of the HCAT project to feed the GMT's NGWS-P for testing the real GMT NGWS' internal HDFS and PyWFS hardware. We passed an external FDR for the final HCAT design on Feb 15, 2022 and we expect to finish the final version of the HCAT testbed by Q3 2022. We will report on more results we obtain in the future with the goal of one day officially retiring the GMT high-risk item of phasing performance.

\subsection* {Acknowledgments}
The HCAT testbed program is supported by a NSF/AURA/GMTO risk-reduction program contract to the University of Arizona (GMT-CON-04535, Task Order No. D3 High Contrast Testbed (HCAT), PI Laird Close). The authors acknowledge support from the NSF Cooperative Support award 2013059 under the AURA sub-award NE0651C. Support for this work was also provided by NASA through the NASA Hubble Fellowship grant \#HST-HF2-51436.001-A awarded by the Space Telescope Science Institute, which is operated by the Association of Universities for Research in Astronomy, Incorporated, under NASA contract NAS5-26555. Alex Hedglen received a University of Arizona Graduate and Professional Student Council Research and Project Grant in February 2020, which helped provide funds for the Holey Mirror for p-HCAT. Alex Hedglen and Laird Close were also partially supported by NASA eXoplanet Research Program (XRP) grants 80NSSC18K0441 and 80NSSC21K0397 and the Arizona TRIF/University of Arizona "student link" program. We are very grateful for support from the NSF MRI Award \#1625441 (for MagAO-X) and funds for the GMagAO-X CoDR from the University of Arizona Space Institute (PI Jared Males) as well.


\bibliography{article}   

\begin{thebibliography}{10}

\bibitem{bryson_2021}
S.~{Bryson}, M.~{Kunimoto}, R.~K. {Kopparapu}, {\em et~al.}, ``{The Occurrence
  of Rocky Habitable-zone Planets around Solar-like Stars from Kepler Data},''
  {\em The Astronomical Journal} {\bf 161}, 36  (2021).

\bibitem{thompson_2018}
S.~E. {Thompson}, J.~L. {Coughlin}, K.~{Hoffman}, {\em et~al.}, ``{Planetary
  Candidates Observed by Kepler. VIII. A Fully Automated Catalog with Measured
  Completeness and Reliability Based on Data Release 25},'' {\em The
  Astrophysical Journal Supplement Series} {\bf 235}, 38  (2018).

\bibitem{guyon_exao}
O.~Guyon, ``Extreme adaptive optics,'' {\em Annual Review of Astronomy and
  Astrophysics} {\bf 56}(1), 315--355  (2018).

\bibitem{males_gmagaox}
J.~{Males}, L.~M. {Close}, O.~{Guyon}, {\em et~al.}, ``{GMagAO-X: extreme
  adaptive optics \& coronagraphy for GMT at first light},'' in {\em Bulletin
  of the American Astronomical Society},   {\bf 51}, 236  (2019).

\bibitem{pacheco}
F.~Quirós-Pacheco, D.~Schwartz, K.~Das, {\em et~al.}, ``{The Giant Magellan
  Telescope phasing strategy and performance},'' in {\em Ground-based and
  Airborne Telescopes VII},  H.~K. Marshall and J.~Spyromilio, Eds.,  {\bf
  10700}, 181 -- 191, International Society for Optics and Photonics, SPIE
  (2018).

\bibitem{schwartz_2017}
N.~Schwartz, J.-F. Sauvage, C.~Correia, {\em et~al.}, ``Sensing and control of
  segmented mirrors with a pyramid wavefront sensor in the presence of
  spiders,'' {\em Proceedings of the Adaptive Optics for Extremely Large
  Telescopes 5}   (2017).

\bibitem{pinna_2006}
E.~Pinna, S.~Esposito, A.~Puglisi, {\em et~al.}, ``{Phase ambiguity solution
  with the Pyramid Phasing Sensor},'' in {\em Ground-based and Airborne
  Telescopes},  L.~M. Stepp, Ed.,  {\bf 6267}, 1028 -- 1037, International
  Society for Optics and Photonics, SPIE  (2006).

\bibitem{pinna_2014}
E.~Pinna, G.~Agapito, F.~Quirós-Pacheco, {\em et~al.}, ``{Design and numerical
  simulations of the GMT Natural Guide star WFS},'' in {\em Adaptive Optics
  Systems IV},  E.~Marchetti, L.~M. Close, and J.-P. Véran, Eds.,  {\bf 9148},
  898 -- 912, International Society for Optics and Photonics, SPIE  (2014).

\bibitem{males_2020}
J.~R. Males, L.~M. Close, O.~Guyon, {\em et~al.}, ``{MagAO-X first light},'' in
  {\em Adaptive Optics Systems VII},  L.~Schreiber, D.~Schmidt, and E.~Vernet,
  Eds.,  {\bf 11448}, 918 -- 925, International Society for Optics and
  Photonics, SPIE  (2020).

\bibitem{close_2018}
L.~M. Close, J.~R. Males, O.~Durney, {\em et~al.}, ``{Optical and mechanical
  design of the extreme AO coronagraphic instrument MagAO-X},'' in {\em
  Adaptive Optics Systems VI},  L.~M. Close, L.~Schreiber, and D.~Schmidt,
  Eds.,  {\bf 10703}, 1227 -- 1236, International Society for Optics and
  Photonics, SPIE  (2018).

\bibitem{haffert_2022}
S.~Y. {Haffert}, L.~M. {Close}, A.~D. {Hedglen}, {\em et~al.}, ``{The
  Holographic Dispersed Fringe Sensors (HDFS): phasing the Giant Magellan
  Telescope},'' {\em arXiv e-prints} , arXiv:2206.03615  (2022).

\bibitem{thomas-osip-2008}
J.~E. Thomas-Osip, G.~Prieto, M.~Johns, {\em et~al.}, ``{Giant Magellan
  Telescope site evaluation and characterization at Las Campanas
  Observatory},'' in {\em Ground-based and Airborne Telescopes II},  L.~M.
  Stepp and R.~Gilmozzi, Eds.,  {\bf 7012}, 757 -- 768, International Society
  for Optics and Photonics, SPIE  (2008).

\bibitem{close_gmagaox}
L.~M. Close, J.~R. Males, A.~Hedglen, {\em et~al.}, ``{Concept for the GMT
  High-Contrast Exoplanet Instrument GMagAO-X and the GMT High-Contrast Phasing
  Testbed with MagAO-X},'' {\em Proceedings of the Adaptive Optics for
  Extremely Large Telescopes 6}   (2019).

\bibitem{astro2020}
E.~National Academies~of Sciences and Medicine, {\em Pathways to Discovery in
  Astronomy and Astrophysics for the 2020s}, The National Academies Press,
  Washington, DC  (2021).

\bibitem{bertrou_2021}
A.~{Bertrou-Cantou}, E.~{Gendron}, G.~{Rousset}, {\em et~al.}, ``{Confusion in
  differential piston measurement with the pyramid wavefront sensor},'' {\em
  Astronomy \& Astrophysics} {\bf 658}, A49  (2022).

\bibitem{cavarroc_2006}
C.~{Cavarroc}, A.~{Boccaletti}, P.~{Baudoz}, {\em et~al.}, ``{Fundamental
  limitations on Earth-like planet detection with extremely large
  telescopes},'' {\em Astronomy \& Astrophysics} {\bf 447}, 397--403  (2006).

\bibitem{por_2018}
E.~H. Por, S.~Y. Haffert, V.~M. Radhakrishnan, {\em et~al.}, ``{High Contrast
  Imaging for Python (HCIPy): an open-source adaptive optics and coronagraph
  simulator},'' in {\em Adaptive Optics Systems VI},  L.~M. Close,
  L.~Schreiber, and D.~Schmidt, Eds.,  {\bf 10703}, 1112 -- 1125, International
  Society for Optics and Photonics, SPIE  (2018).

\bibitem{melt_2018}
T.~Pfrommer, S.~Lewis, J.~Kosmalski, {\em et~al.}, ``{MELT: an optomechanical
  emulation testbench for ELT wavefront control and phasing strategy},'' in
  {\em Ground-based and Airborne Telescopes VII},  H.~K. Marshall and
  J.~Spyromilio, Eds.,  {\bf 10700}, 1077 -- 1090, International Society for
  Optics and Photonics, SPIE  (2018).

\bibitem{ape_2008}
F.~{Gonte}, C.~{Araujo}, R.~{Bourtembourg}, {\em et~al.}, ``{Active Phasing
  Experiment: preliminary results and prospects},'' in {\em Ground-based and
  Airborne Telescopes II},  L.~M. {Stepp} and R.~{Gilmozzi}, Eds., {\em Society
  of Photo-Optical Instrumentation Engineers (SPIE) Conference Series} {\bf
  7012}, 70120Z  (2008).

\bibitem{vigan_2011}
A.~Vigan, K.~Dohlen, and S.~Mazzanti, ``On-sky multiwavelength phasing of
  segmented telescopes with the zernike phase contrast sensor,'' {\em Appl.
  Opt.} {\bf 50}, 2708--2718  (2011).

\bibitem{chanan_1998}
G.~Chanan, M.~Troy, F.~Dekens, {\em et~al.}, ``Phasing the mirror segments of
  the keck telescopes: the broadbandphasing algorithm,'' {\em Appl. Opt.} {\bf
  37}, 140--155  (1998).

\bibitem{chanan_2000}
G.~Chanan, C.~Ohara, and M.~Troy, ``Phasing the mirror segments of the keck
  telescopes ii: the narrow-band phasing algorithm,'' {\em Appl. Opt.} {\bf
  39}, 4706--4714  (2000).

\bibitem{bouchez_2012}
A.~H. Bouchez, D.~S. Acton, G.~Agapito, {\em et~al.}, ``{The Giant Magellan
  Telescope adaptive optics program},'' in {\em Adaptive Optics Systems III},
  B.~L. Ellerbroek, E.~Marchetti, and J.-P. Véran, Eds.,  {\bf 8447}, 571 --
  582, International Society for Optics and Photonics, SPIE  (2012).

\bibitem{van-dam-2016}
M.~A. van Dam, B.~A. McLeod, and A.~H. Bouchez, ``Dispersed fringe sensor for
  the giant magellan telescope,'' {\em Appl. Opt.} {\bf 55}, 539--547  (2016).

\bibitem{kopon-2016}
D.~Kopon, B.~McLeod, M.~A. van Dam, {\em et~al.}, ``{On-sky demonstration of
  the GMT dispersed fringe phasing sensor prototype on the Magellan
  Telescope},'' in {\em Adaptive Optics Systems V},  E.~Marchetti, L.~M. Close,
  and J.-P. Véran, Eds.,  {\bf 9909}, 1256 -- 1266, International Society for
  Optics and Photonics, SPIE  (2016).

\bibitem{males_2018}
J.~R. Males, L.~M. Close, K.~Miller, {\em et~al.}, ``{MagAO-X: project status
  and first laboratory results},'' in {\em Adaptive Optics Systems VI},  L.~M.
  Close, L.~Schreiber, and D.~Schmidt, Eds.,  {\bf 10703}, 76 -- 89,
  International Society for Optics and Photonics, SPIE  (2018).

\bibitem{close_2020}
L.~M. Close, J.~Males, J.~D. Long, {\em et~al.}, ``{Prediction of the planet
  yield of the MaxProtoPlanetS high-contrast survey for H-alpha protoplanets
  with MagAO-X based on first light contrasts},'' in {\em Adaptive Optics
  Systems VII},  L.~Schreiber, D.~Schmidt, and E.~Vernet, Eds.,  {\bf 11448},
  157 -- 174, International Society for Optics and Photonics, SPIE  (2020).

\bibitem{van-gorkom}
K.~V. Gorkom, K.~L. Miller, J.~R. Males, {\em et~al.}, ``{Characterization of
  deformable mirrors for the MagAO-X project},'' in {\em Adaptive Optics
  Systems VI},  L.~M. Close, L.~Schreiber, and D.~Schmidt, Eds.,  {\bf 10703},
  1266 -- 1272, International Society for Optics and Photonics, SPIE  (2018).

\bibitem{esposito_2000}
S.~Esposito, O.~Feeney, and A.~Riccardi, ``{Laboratory test of a pyramid
  wavefront sensor},'' in {\em Adaptive Optical Systems Technology},  P.~L.
  Wizinowich, Ed.,  {\bf 4007}, 416 -- 422, International Society for Optics
  and Photonics, SPIE  (2000).

\bibitem{van-dam}
M.~A. van Dam, R.~Conan, A.~H. Bouchez, {\em et~al.}, ``{Design of a truth
  sensor for the GMT laser tomography adaptive optics system},'' in {\em
  Adaptive Optics Systems III},  B.~L. Ellerbroek, E.~Marchetti, and J.-P.
  Véran, Eds.,  {\bf 8447}, 482 -- 492, International Society for Optics and
  Photonics, SPIE  (2012).

\bibitem{breckinridge_2015}
J.~B. Breckinridge, W.~S.~T. Lam, and R.~A. Chipman, ``Polarization aberrations
  in astronomical telescopes: The point spread function,'' {\em Publications of
  the Astronomical Society of the Pacific} {\bf 127}, 445--468  (2015).

\bibitem{lam_2015}
W.~S.~T. Lam and R.~Chipman, ``Balancing polarization aberrations in crossed
  fold mirrors,'' {\em Appl. Opt.} {\bf 54}, 3236--3245  (2015).

\end{thebibliography}
\bibliographystyle{spiejour}   


\vspace{2ex}\noindent\textbf{Alexander D. Hedglen} is an Optical Sciences PhD student at the University of Arizona. He received his B.A. in Astronomy and B.S. in Physics from the University of Hawai`i at Hilo in 2016. His research focuses on adaptive optics for the direct imaging of exoplanets.

\vspace{2ex}\noindent\textbf{Sebastiaan Y. Haffert} is a NASA Hubble Postdoctoral Fellow at the University of Arizona’s Steward Observatory. His research focuses on high-spatial and high-spectral resolution instrumentation
for exoplanet characterization.

\vspace{1ex}
\noindent Biographies and photographs of the other authors are not available.

\listoffigures
\listoftables

\end{spacing}
\end{document}